\documentclass[a4paper,onecolumn,11pt,accepted=2023-02-23]{quantumarticle}
\pdfoutput=1
\usepackage[utf8]{inputenc}
\usepackage[english]{babel}
\usepackage[T1]{fontenc}
\usepackage{amsmath}
\usepackage[colorlinks]{hyperref}

\usepackage{tikz}
\usepackage{lipsum}

\usepackage{enumitem}
\usepackage{soul}
\usepackage{mathtools}
\usepackage{amsfonts}
\usepackage{geometry}
\usepackage{multicol}
\usepackage{cite}
\usepackage{wrapfig}
\usepackage{setspace}
\usepackage{titlesec}

\usepackage[nice]{nicefrac}
\usepackage{array}
\usepackage{graphicx}
\usepackage{dcolumn}
\usepackage{amsthm,amssymb,color}
\usepackage{bbm}
\usepackage{bm}
\usepackage{synttree}
\usepackage{caption}
\usepackage{subfig}

\newcommand{\COMMENT}[1]{}

\newtheorem*{conjecture*}{Conjecture}
\theoremstyle{definition}
\newtheorem{definition}{Definition}[section]
\newtheorem{example}[definition]{Example}

\usepackage[numbers,sort&compress]{natbib}

\begin{document}

\title[]{A mathematical framework for operational fine tunings} 

\author{Lorenzo Catani} 
\ead{lorenzo.catani4@gmail.com}
\author{Matthew Leifer}
\address{Institute for Quantum Studies \& Schmid College of Science and Technology, Chapman University, One University Drive, Orange, CA, 92866, USA} 

\begin{abstract}
In the framework of ontological models, the inherently nonclassical features of quantum theory always seem to involve properties that are fine tuned, \emph{i.e.}\ properties that hold at the operational level but break at the ontological level.  Their appearance at the operational level is due to unexplained special choices of the ontological parameters, which is what we mean by a fine tuning.  Famous examples of such features are contextuality and nonlocality. In this article, we develop a theory-independent mathematical framework for characterizing \emph{operational} fine tunings. These are distinct from \emph{causal} fine tunings -- already introduced by Wood and Spekkens in [NJP,$\bold{17}$ 033002(2015)] -- as the definition of an operational fine tuning does not involve any assumptions about the underlying causal structure.
We show how known examples of operational fine tunings, such as Spekkens' generalized contextuality, violation of parameter independence in Bell experiment, and ontological time asymmetry, fit into our framework.  We discuss the possibility of finding new fine tunings and we use the framework to shed new light on the relation between nonlocality and  generalized contextuality.  Although nonlocality has often been argued to be a form of contextuality, this is only true when nonlocality consists of a violation of parameter independence.
We formulate our framework also in the language of category theory using the concept of functors.

\end{abstract}

\maketitle

\section{Introduction}


Although quantum theory is almost a century old, there is still no consensus on what it says about the nature of reality.  Multiple interpretations are on the table, each giving a radically different account of the world \cite{Everett1957,Wallace2012,Bohm1952,Durr2009,Ghirardi1986,Bassi2013,Rovelli1996,Modalinterpretations,Brukner2003,Pitowsky2006,Fuchs2014}.  It is our, surprisingly controversial, belief that there is a correct understanding of quantum theory to be had, but that none of the existing interpretations are up to the task.

One reason for believing this is that many of the oft remarked novelties of quantum theory---such as superposition, interference, entanglement, and no cloning---exist in classical phase space models in which there is a restriction on what can be known, such as Spekkens' toy theory \cite{Spekkens2007, Spekkens2016, CataniBrowne2017,Catani2021}.  Therefore, these phenomena have perfectly straightforward and reasonable explanations.  In contrast, most interpretations of quantum theory explain these phenomena in radically nonclassical terms.  For example, the account of the Einstein-Podolsky-Rosen experiment given by the de Broglie-Bohm theory \cite{Norsen2017} involves nonlocal influences, even though the correlations in this experiment have a perfectly natural local model.  It is only with more sophisticated Bell-type experiments that nonlocality is actually required.  To give another example, the Everett/many-worlds interpretation implies that the universe is constantly splitting into multiple copies whenever apparently probabilistic processes occur \cite{Wallace2012}, even when those processes are well accounted for by classical statistical mechanics.

If a quantum phenomenon has a natural classical model, then that phenomenon cannot be the source of an advantage in quantum information processing, as the classical model could be run as a simulation on a classical device.  So interpretations that deploy highly nonclassical explanations liberally, even in those cases where none are required, can provide no guide to where quantum advantage is likely to be found.  In our view, one of the things that characterizes a scientific truth (over and above other types of truth that may exist) is that it offers pragmatic value.  So the correct interpretation of quantum theory should also be the most useful.  With this in mind, we believe that a good interpretation should minimize quantum weirdness, offering straightforward explanations where they are adequate, and more subtle explanations only where they are needed. 

With these remarks, we do not wish to downplay the achievements of existing interpretations.  In the early days, it was difficult enough to come up with accounts of quantum theory that were logically coherent, consistent with the empirical facts, and that deal adequately with the problems surrounding quantum measurement.  However, having achieved this aim, we ought to move beyond consistency and towards correctness.  We need to put forward criteria that a correct account ought to satisfy, and identify where current interpretations fall short of them. 

To date, the quantum phenomena that are most universally viewed as problematic are those arising from the no-go theorems of Bell \cite{Bell} and Kochen-Specker \cite{KochenSpecker,Spekkens2005}, namely nonlocality and contextuality.  A related one, which has recently been discovered, is the breaking of time symmetry \cite{Price2012, LeiferPusey}.  Dozens of other no-go theorems have been derived, but their significance is more controversial \cite{LeiferOntology}.  In light of this, we should ask: what makes a good assumption for a no-go theorem?  Which assumptions provide significant, not overly strong constraints that we ought to consider in our theory-building? 

In recent years, an answer to this question has begun to emerge.  Good assumptions are those that assert that there is no \emph{fine tuning} in the theory that underlies quantum theory.\footnote{It is worth mentioning that the first contribution that introduced the notion of fine tuning in this context is due to Valentini \cite{Valentini1996}. He highlighted how the conflict between the fundamental nonlocality of quantum mechanics and the operational validity of Einstein’s relativity is resolved at the price of introducing what we here call a fine tuning. In particular, this motivated him to develop his version of Bohmian mechanics \cite{Bohm1952}, where the ``fine tuned'' property of no-signalling emerges from an equilibration process, while not holding at the fundamental level.}  One of the main goals of this work is to build a mathematically precise theory of this concept. 
\footnote{Notice that the requirement of no fine tuning that we describe here is stronger than the requirement of no fine tuning as ``naturalness'' used in cosmology and particle physics \cite{Weinberg1989, Williams2015}. The latter aims to prevent that the parameters of a certain model must be adjusted very precisely in order to fit with certain empirical observations. In our case we demand that the parameters of any underlying physical theory should not be adjusted very precisely in order for its predicted statistics to fit with the properties that the operational theory stipulates to exist \textit{in principle}, \textit{i.e.} according to the predictions provided by the mathematical formulation of the theory. In other words, the no fine tuning we consider does not just require for some parameters of a \textit{particular} model to fit with some particular observations (\textit{e.g.} to match the expected value of the cosmological constant), but for the parameters of \textit{any} underlying physical theory (any ``ontic extension'' in the jargon that we will use later) to provide statistics that fit with the universal properties predicted by the operational theory.}
 


Fine tuning based theorems work as follows.  We first identify a property of the operational predictions of the theory that always holds.  We assume that the analogous property should also hold at the ontological level in whatever underlying theory describes the true physics of the system.  Then, under appropriate additional assumptions, we show that this leads to a contradiction---the property cannot hold at the ontological level.

An example may be helpful here.  Consider two preparation procedures for a physical system that always yield the same outcome probabilities for any measurement that we can perform on the system.  We call such preparations \emph{operationally equivalent}.  It is natural to assume that the operational equivalence holds because the two preparations yield the same distribution of properties in the underlying physical theory.  After all, the simplest explanation of why two things look exactly the same is that they are, in fact, the same.  This assumption is known as \emph{preparation noncontextuality}, and was introduced by Spekkens \cite{Spekkens2005} who argues that it is a form of Leibniz principle of the identity of indiscernibles \cite{Spekkens2019}.  Spekkens also showed that a preparation noncontextual model for quantum theory is impossible \cite{Spekkens2005}, which we summarize by saying that quantum theory is \emph{preparation contextual}.

If two preparation procedures yield different distributions of physical properties, one would naively expect to be able to detect the difference between them. However, nature does not give us direct access to the physical properties of a system.  The only information that we can access is the results of quantum measurements performed on the system, and these measurements might only yield coarse-grained information about the true underlying physical state.  If the coarse-graining is set up in precisely the right way, the physical differences between the two preparation procedures can be impossible to detect, rendering the preparations operationally equivalent.

The question then becomes why the coarse-graining happens to be set up in precisely the right way.  It requires the parameters of the model to have specially chosen values, and there is no law of physics that explains why they are chosen this way.  The model is fine tuned. Notice that such fine tuning is of problematic interpretation because it characterizes the model with a conspiratorial connotation: the differences between certain preparations present in the model are hidden in the operational theory without a physical explanation within the model. We will discuss the possible solutions to the fine tuning problem in section \ref{Conclusion}.

Preparation contextuality is an example of an \emph{operational} fine tuning, which is distinct from the notion of a \emph{causal} fine tuning in the framework of classical causal models \cite{Pearl}.  The causal models framework works with causal structures specified by a directed acyclic graph (DAG).  Each node in the graph represents an observable variable and the structure of the DAG imposes conditional independence relations between the variables, known as the causal Markov condition.  Given a set of variables and a probability distribution over them, one can determine which DAGs are compatible with the probabilities in the sense that the probability distribution obeys all the conditional independence relations implied by the DAG.  If, for all compatible DAGs, the probability distribution obeys additional conditional independence relations that are not implied by the DAG then the probability distribution has a causal fine tuning.  This is because, if the parameters of the causal model were varied slightly then these additional conditional independences would fail to hold.  Thus, it takes a special choice of parameters to have these conditional independences, and no causal explanation is available for them.  The assumption that a model does not have a causal fine tuning is called \emph{faithfulness} in the literature on causal models.

In 2015, Wood and Spekkens \cite{Wood2015} introduced the notion of causal fine tunings into quantum foundations.  They showed that all causal models compatible with the probabilities that violate a Bell inequality are causally fine tuned.  In contrast, this work deals with \emph{operational} fine tunings, which can be defined without any reference to the underlying causal structure\footnote{Additional causal assumptions are usually needed to derive a no-go theorem about these fine tunings, but the point is that these fine tunings do not require causal assumptions for their definition.}. As in the example of preparation contextuality, operational fine tunings are defined in terms of equivalences between statistics of different experiments and the requirement that such equivalences must be preserved at the ontological level. To define operational fine tunings, we do not need to assume the full ontological models framework \cite{Harrigan}, but the weaker notion of an \emph{ontic extension} \cite{LeiferPusey} that just encodes a simple form of realism, without the accompanying causal assumptions that are built into ontological models.
 
The requirement of no operational fine tunings can be seen as a requirement of structure preservation between the operational and the ontological level. The branch of mathematics that deals with structure preservation between different realms is category theory \cite{Leinster2014}, and we reformulate our framework in categorical terms in section \ref{Cats}.  More precisely, we define operational and ontological categories. The former describes all the possible statistics associated with the experiments under consideration, while the latter refers to the corresponding ontological representations. An operational theory, like quantum mechanics, puts restrictions on the possible experimental statistics and it is described by a subcategory of the operational category.  An ontic extension of the operational theory is a functor from the operational theory to the ontological category.  An operational property, represented by an equation in the operational theory, is not fine tuned in an ontic extension if the ontic extension maps the equation to the analogous one in the ontological category. 

In summary, we provide a rigorous mathematical framework, developed also in the language of category theory and functors, that characterizes operational fine tunings. 
In addition to accounting for all the known operational fine tunings -- associated to violations of generalized noncontextuality, parameter independence and time symmetry -- the framework describes more general ones, thus setting the ground for formulations of further no-go theorems.
In light of our framework and the distinction we draw between operational and causal fine tunings, we analyze the notion of Bell's local causality, that can be conceived in two ways. On the one hand, in the classical causal model framework, it can be seen as a requirement of no causal fine tuning \cite{Wood2015}; on the other hand, it can be decomposed into the assumptions of parameter independence -- justified by a requirement of no operational fine tuning --  and outcome independence -- an assumption of purely causal nature \cite{Jarrett1984}. Through this consideration we deepen the understanding of the relation between nonlocality and generalized contextuality, where the former is not just an example of the latter, as usually stated,  because it can be obtained, unlike contextuality, by involving a purely causal fine tuning.

The remainder of this article is structured as follows. We start, in section \ref{Causality}, by specifying the approach on causality taken in this work. We define the operational framework in section \ref{Operational} and the ontological framework in section \ref{Ontological}. We report the already known examples of operational fine tunings in section \ref{Examples}. We construct our generalization of operational fine tunings, first introducing the notion of \textit{classical processings} in section \ref{Proc}, and then formulating the no fine tuning requirement in section \ref{Generalization}. We provide the categorical formulation in section \ref{Cats}. We discuss the future avenues and the applications of the framework, \emph{e.g.} possible new fine tunings, and the analysis of the relation between nonlocality and contextuality, in section \ref{Conclusion}.

\section{A word on causality}
\label{Causality}

The framework of causal modeling, as developed by Pearl et.\ al.\  \cite{Pearl}, is becoming increasingly popular in quantum foundations \cite{Wood2015,Ried2015,Chaves2015,Fritz2016,Costa2016,Allen2017,Weilenmann2017,Wolfe2019,Vilasini2019,Weilenmann2020,Barrett2020}.  Such \textit{Pearl-causal} approach treats causality as fundamental (or at least unanalyzed).  The primitive notion of the approach is the one of causal structure, that places constraints on what can be observed operationally, and on what can happen in an underlying realist model, which is usually treated as a classical causal model.  No-go theorems reveal fine tunings in those classical models.  This is in line with the Wood-Spekkens treatment of Bell's theorem \cite{Wood2015} and has been developed into a more general framework in \cite{Cavalcanti2018}.

Part of the motivation for this approach is the idea that the most important part of a realist explanation is that it is causal.  By replacing a classical causal model with some more general notion of a quantum causal model, one might be able to evade the no-go theorems while salvaging the notion of a causal explanation \cite{Allen2017,Barrett2020}.
This causally-centered approach is quite different from the usual approach to realist theories in quantum foundations in which, to count as realist, a theory must specify an ontology (the things that exist, independent of observers) as well as the laws (rules for how the ontology behaves).  The laws will typically give rise to causal explanations, \textit{i.e.} the state of affairs at one time can be used to determine the state of affairs at a later time using the laws.  Causality is a consequence of the laws in the standard realist approach, but causality is not, in itself, fundamental.   In contrast, the Pearl-causal approach is willing to jettison much of this structure and takes the goal of providing causal explanations as fundamental.  It can be thought as an approach which lies at half-way between operational approaches, which jettison the idea of causal explanation entirely, and standard realist approaches. 

While there is a lot to admire in the Pearl-causal approach, we are not yet ready to take causal structure as fundamental in physics, and we do not do so in this work.  From statistical mechanics, there are good reasons to believe that causality is not fundamental in physics, but rather emergent from the conditions that give rise to the second law of thermodynamics.  From the practical perspective, employing fundamental causality would make it difficult to formulate the time-symmetry fine tuning \cite{LeiferPusey}, which relates experiments with oppositely directed arrows of time.  Even more seriously, the Pearl-causal approach rules out the possibility that the ontological description may have no notion of causality, and that causality is emergent at the operational level.  This might lead to a different perspective on how seriously we need to take the causal fine tunings identified by Wood and Spekkens.

Although we do not take causality as fundamental, we cannot eliminate causal notions from our framework entirely.  This is because the only part of quantum theory that is universally agreed upon is its operational predictions. 
There is no agreement on which elements of the quantum formalism, if any, represent reality, but everyone agrees with the experimental statistics that it predicts, \textit{i.e.} the probabilities of outcomes given certain experimental procedures (specified by a list of instructions of what to do in the lab). 
Therefore, when investigating the general question of what the possible theories underlying quantum theory must look like, we have to start with the operational description.  This is by no means to say that we must be operationalists, but just that we have to say how the agreed upon operational predictions are accounted for by the underlying theory.  If we are investigating a specific proposed underlying theory, such as the de Broglie-Bohm theory, then we can just deal with the theory in its own terms and explain how the operational predictions emerge from it.  This is the correct order of explanation for a realist theory.  However, if we are trying to derive constraints on \emph{all possible} realist theories then we cannot work in this way, as different theories may account for the predictions in very different ways.  So, we have to start with the operational theory and work backwards from it.

This creates a problem for eliminating causality because operational frameworks necessarily have some causal notions built into them.  The goal of an operational framework is to predict what will happen when the experimenter performs certain actions.  There are variables that the experimenter can control, such as the settings of knobs and switches on their equipment, and variables that they can only observe, such as the measurement readouts.  But the distinction between variables that are controllable and those that are only observable is a causal notion, dependent on the arrow of time.  To see this, consider a switch that can be in one of two positions, labeled $0$ and $1$.  Part of the specification of an experimental procedure might be ``set the switch to zero''.  To do this, the experimenter has to be able to set the switch to $0$ regardless of whether it is initially set to $0$ or $1$.  If we imagine that there is a 50/50 chance of the switch being initially set to $0$ or $1$ then the experimenter has to erase a bit of information in order to set the switch to $0$.  Landauer's principle \cite{Landauer1961} tells us that this has a thermodynamics cost -- the entropy of the environment surrounding the switch must increase.  This is so even if there is not a 50/50 chance of the two switch settings.  Providing there is a nonzero chance that the switch is initially set to $1$, setting it to zero erases a nonzero amount of information.  Therefore, the experimenter can only control the switch setting if there is a supply of low entropy systems in the environment to dump the entropy into.  The controllability of the switch is not fundamental, but emergent from the thermodynamic arrow of time.  If the universe were in a state of global thermal equilibrium then the experimenter would only be able to observe, but not control, the switch setting.\footnote{To say nothing of the fact that experimenters and switches with metastable states would not exist in such a universe.} Similarly, an apparently uncontrollable variable might become controllable if the arrow of time were different.  What is an uncontrollable measurement outcome to the experimenter might be controllable for a hypothetical experimenter with a reversed arrow of time.

For this reason, we think that the distinction between controllable and observed variables ought to play no role in a fundamental physical theory, but we are stuck with the fact that the only universally agreed upon part of quantum theory is formulated in terms of this distinction.  

\section{The operational framework}

\label{Operational}

This section introduces the operational framework that we use to describe physical experiments.  Our main interest lies in how these experiments must be represented at the ontological level, \textit{i.e.} what they say about reality, but we need an operational description to start with, especially if we want the framework to apply more broadly than just to classical and quantum theories. The corresponding ontological framework is developed in section \ref{Ontological}.

In our operational framework, an \emph{experiment} $E$ is composed of three things: the variables involved in the experiment, a causal structure specifying how those variables are arranged in space-time 
and a probability distribution over the variables predicted by the theory under consideration.  We discuss each of these in turn. 

In an experiment, there are some variables that are controlled by the experimenter, and some variables that they observe but do not directly control.  A controlled variable may refer to the choice of setting of an experimental apparatus, such as turning a knob that tunes the frequency of a laser or selecting the orientation of a Stern-Gerlach magnet, while observed variables,  that are not under the control of the experimenter, might record whether or not a detector clicks or the reading on a measurement device.

Let $\bm{C} = \{C_1,C_2,\cdots,C_m\}$ be the set of controlled variables in an experiment and $\bm{O} = \{O_1,O_2,\cdots,O_n\}$ be the set of observed variables.  These two sets are always disjoint.  As indicated, we assume both of these sets are finite for simplicity.  Let $\bm{X} = \bm{C} \cup \bm{O}$ be the set of all variables involved in the experiment.

Each variable $X \in \bm{X}$ has a set of possible values $v(X)$.  For example, if $X$ is a controlled variable representing the tuning frequency of a laser then $v(X)$ would be the set of possible frequencies that the laser can be tuned to.  For simplicity, we assume that $v(X)$ is a finite set.  We use lower case letters to denote elements of these sets, so if $X \in \bm{X}$ then $X = x$ means that $X$ takes the value $x$ for some $x \in v(X)$.

For a set of variables $\bm{Y}$, which could be $\bm{C}$, $\bm{O}$ or $\bm{X}$, we define \[v(\bm{Y}) = \bigtimes_{Y\in \bm{Y}} v(Y),\] where the symbol  $\bigtimes$ denotes the Cartesian product.
If $\bm{Y} = \{Y_1,Y_2,\cdots, Y_n\}$ and $\bm{y} \in v(\bm{Y})$ then the expression $\bm{Y} = \bm{y}$ should be read as $Y_1 =y_1, Y_2 = y_2, \cdots Y_n = y_n$, where $y_j \in v(Y_j)$ is the appropriate component of $\bm{y}$.  The comma here denotes a logical AND, so this should be read as $Y_1$ takes the value $y_1$ AND $Y_2$ takes the value $y_2$ AND $\ldots$ AND $Y_n$ takes the value $y_n$.

\begin{figure*}[!htb]
	\centering
	{\includegraphics[scale=0.42]{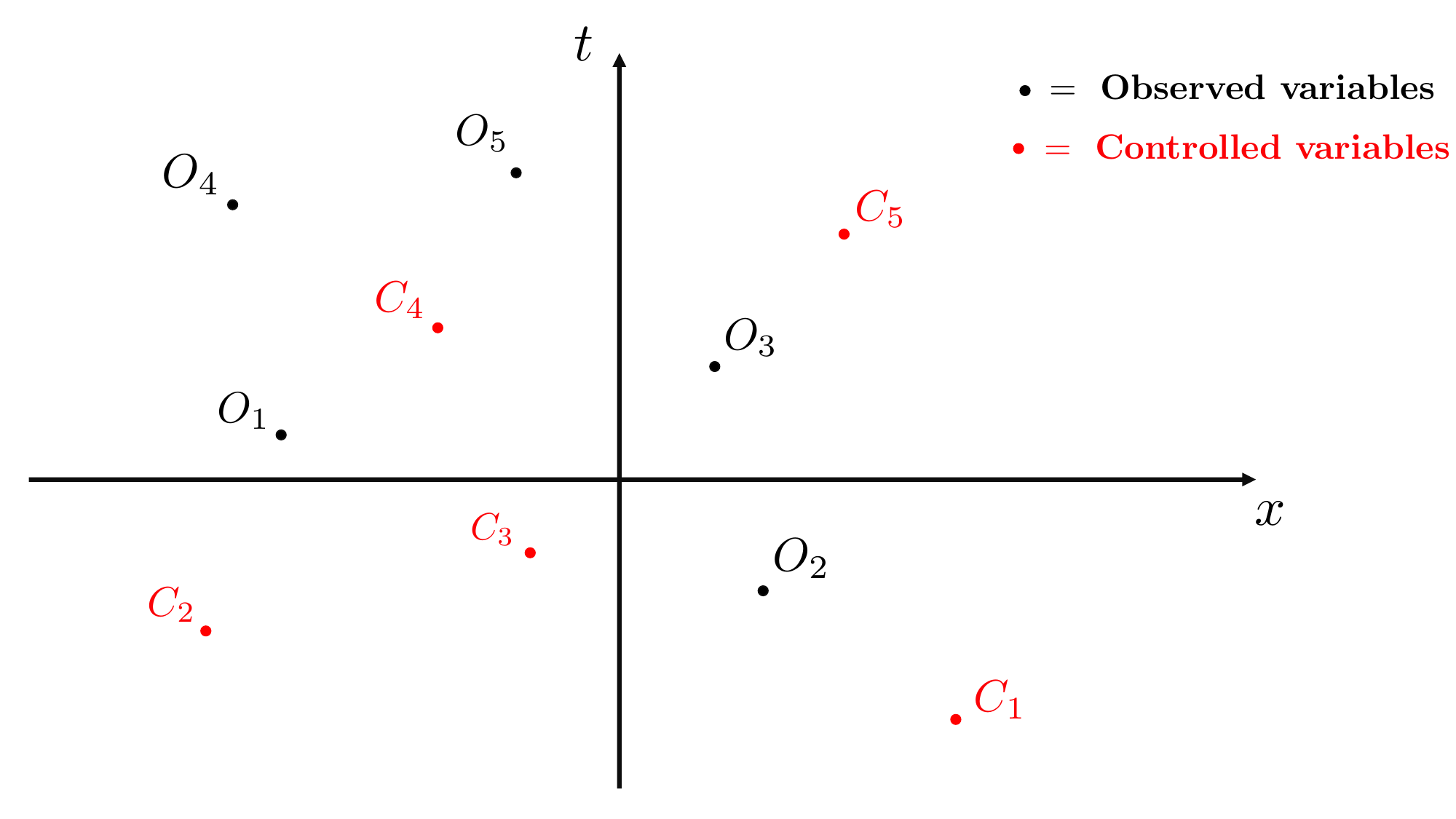}}
	\caption{A generic experimental scenario. The operational variables can be divided into those that are controlled by the experimenter (in red) and those that are only observed (in black).}
	\label{ExperimentA}
\end{figure*}

On a given run of the experiment, each variable $X \in \bm{X}$ is associated with a small localized region of space-time---the region in which the experimental setting is made or the observation is first recorded.  These form a pattern in space-time that we assume can be replicated in disjoint regions of space-time such that they constitute independent runs of the experiment.  Such a pattern is illustrated in figure~\ref{ExperimentA}.

\vspace{5mm}

\newpage
In principle, the full specification of the pattern in space-time might be an important part of the experiment, but, for the examples we wish to consider, only the causal structure is important.  Thus, we will assume that $\bm{X}$ is equipped with a preorder relation \footnote{Let us recall that a preorder is an homogeneous relation (a subset of the Cartesian product), here denoted with $\preceq$, that satisfies the properties of \textit{reflexivity}, \textit{i.e.} $X\preceq X$ $\;\forall \;X\in \bm{X}$ and \textit{transitivity}, \textit{i.e.} $X_1\preceq X_2$, $X_2\preceq X_3$ implies $X_1\preceq X_3$ $\;\forall \; X_1,X_2,X_3 \in \bm{X}$. If the preorder satisfies also \textit{antisymmetry},  \textit{i.e.} $X_1\preceq X_2$ and $X_2\preceq X_1$ imply $X_1=X_2$, then it is called \textit{partial order}. Here we do not assume that $\preceq$ is a partial order.  The regions of space-time occupied by the variables have nonzero extent, \textit{i.e.} they are not points.  For example, the digital readout on a measuring device occupies a nonzero amount of space and it takes a finite amount of time for a reading to fully appear on it.  Because of this, in principle, more than one variable might be associated with the same localized region, so we do not want the relation to be antisymmetric.} $\preceq$, where $X \preceq X'$ means that $X'$ is in the future lightcone of $X$. \footnote{In studies of causal fine tunings it is more common to specify the causal structure by a directed acyclic graph (DAG) rather than a preorder.  The reachability relation of a DAG is a partial order, \textit{i.e.} a preorder which is antisymmetric. For this reason, the preorder is less specific than a DAG. Unlike a DAG, a preorder also allows more than one variable to occupy the same region.  Because we are developing a framework for \emph{operational} rather than \emph{causal} fine tunings, we want to minimize the amount of causal structure we impose.} 
We use $X \nsim X'$ to mean that neither $X \preceq X'$ nor $X' \preceq X$, which means that $X$ and $X'$ are space-like related.  We will specify the causal arrangement of an experiment by just giving the preorder, rather than a detailed pattern in space-time.


Finally, an operational theory should make predictions for the outcomes of an experiment, so we assume that the experiment $E$ specifies a conditional probability distribution $p_E(\bm{O} = \bm{o}|\bm{C} = \bm{c})$ for the observed variables given the controlled variables, where $\bm{o} \in v(\bm{O})$ and $\bm{c} \in v(\bm{C})$.  To simplify the notation, we often write this as $p_E(\bm{O}|\bm{C})$.  An equation involving $p_E(\bm{O}|\bm{C})$ is assumed to hold for every possible choice of value assignments, e.g.\ $p_E(\bm{O}|\bm{C}) = p_{E'}(\bm{O}'|\bm{C}')$ means that $v(\bm{O}) = v(\bm{O}')$, $v(\bm{C}) = v(\bm{C}')$, and $p_E(\bm{O} = \bm{o}|\bm{C}=\bm{c}) = p_{E'}(\bm{O}' = \bm{o}|\bm{C}'=\bm{c})$ for all $\bm{o}\in v(\bm{O})$ and $\bm{c} \in v(\bm{C})$.  Similarly $\sum_{\bm{O}} p_E(\bm{O}|\bm{C})$ should be understood as $\sum_{\bm{o} \in v(\bm{O})} p_E(\bm{O} = \bm{o}|\bm{C})$.  We will often omit the subscript $_E$ on $p_E(\bm{O} = \bm{o}|\bm{C})$ when the experiment is clear from context.

In summary, an experiment is defined as follows.

\begin{definition}
	An \emph{experiment} is a structure $E = (\bm{C},\bm{O},\preceq,v,p)$, where
	\begin{itemize}
		\item $\bm{C}$ is the set of \emph{controlled variables}.
		\item $\bm{O}$ is the set of \emph{observed variables}.
		\item $\bm{C}$ and $\bm{O}$ are disjoint.
		\item $\preceq$ is a preorder on $\bm{C} \cup \bm{O}$ specifying the causal structure of the experiment.
		\item $v$ is a map from $\bm{C} \cup \bm{O}$ to the set of finite sets.  $v(X)$ is the set of \emph{values} that the variable $X$ can take.
		\item $p$ is a conditional probability distribution for the observed variables given the controlled variables, which we also denote $p_E(\bm{O}|\bm{C})$. 
	\end{itemize}
\end{definition}

\section{The ontological framework}

\label{Ontological}

A natural way of explaining the predictions of an experiment $E$ is to assume that the experiment probes the properties of actual physical systems that exist independently of the experimenter. We call these physical properties \emph{ontic states}.  The variable $\Lambda$ is used to describe the ontic state that the system occupies, with possible values -- the ontic states -- denoted by $\lambda$.  The set of possible ontic states is called the \emph{ontic state space} and is denoted $v(\Lambda)$.  We assume that $v(\Lambda)$ is a measurable space because we need to assign probabilities to the ontic states. 

We further impose the following assumptions.
\begin{enumerate}

\item \emph{(Single world) Realism}. The operational and ontic variables each take a definite value on each run of the experiment.
\item  \emph{Independence}. All the runs of the experiment are independent and identically distributed. 
\item  \emph{Free choice}. The experimenter may choose the controlled variables $\bm{C}$ however they like, independently of $\bm{O}$ and $\Lambda$.  This will usually be done by the experimenter specifying a probability distribution $p(\bm{C})$ over the controlled variables.
\end{enumerate}

Taken together, these assumptions imply the existence of a joint conditional probability distribution associated to the operational and ontic variables, $q_E(\bm{O}, \Lambda |\bm{C}).$ The third assumption implies that we can always work with distributions that are conditioned on $\bm{C}$, without worrying that $\bm{C}$ might be influenced by the other variables.

\begin{definition}
An \emph{ontic extension} of an experiment $E$ consists of an ontic state space $v(\Lambda)$ for the ontic variable $\Lambda$ and a conditional probability distribution $q_E(\bm{O},\Lambda|\bm{C})$ that reproduces the operational predictions, in the sense that
\begin{equation}
	\label{eq:Reprod}
	p_E(\bm{O} |\bm{C})=\sum_{\Lambda} q_E(\bm{O}, \Lambda | \bm{C}).
\end{equation}
\end{definition}
Note that the sum over ontic states $\Lambda$ should be replaced by an integral if the ontic state space is continuous.  With this understood, we will work with sums in what follows.  Note also that we always use $p$ to denote the probability distributions in an operational theory and $q$ to denote the corresponding distributions in an ontic extension, so equation~\eqref{eq:Reprod} can be summarized as $q_E(\bm{O}|\bm{C}) = p_E(\bm{O}|\bm{C})$.

The notion of an ontic extension is more general than the more popular notion of an ontological model \cite{Harrigan}.  It is so general that every experiment admits a trivial ontic extension with $q_E(\bm{O},\Lambda|\bm{C}) = p_E(\bm{O}|\bm{C})q(\Lambda)$, where $q(\Lambda)$ is any fixed probability distribution over $v(\Lambda)$, and $v(\Lambda)$ is any measurable set.  This does not qualify as a model in which the ontic state explains the operational observations in any way, and there is no question of deriving a no-go theorem for such a model, so we typically do impose additional assumptions to rule out these trivial models.  However, the \emph{definition} of an operational fine tuning can be made at this level of generality.

The additional assumptions that go into the ontological models framework are causal in nature, \textit{i.e.} we would make assumptions about how $\Lambda$ fits into the causal preorder of the experiment and impose conditions on the probability distribution based on the preorder.  While the full details of this are not relevant to the present work, it is helpful to consider a specific example. 

\begin{figure*}[!htb]
	\centering
	{\includegraphics[scale=0.36]{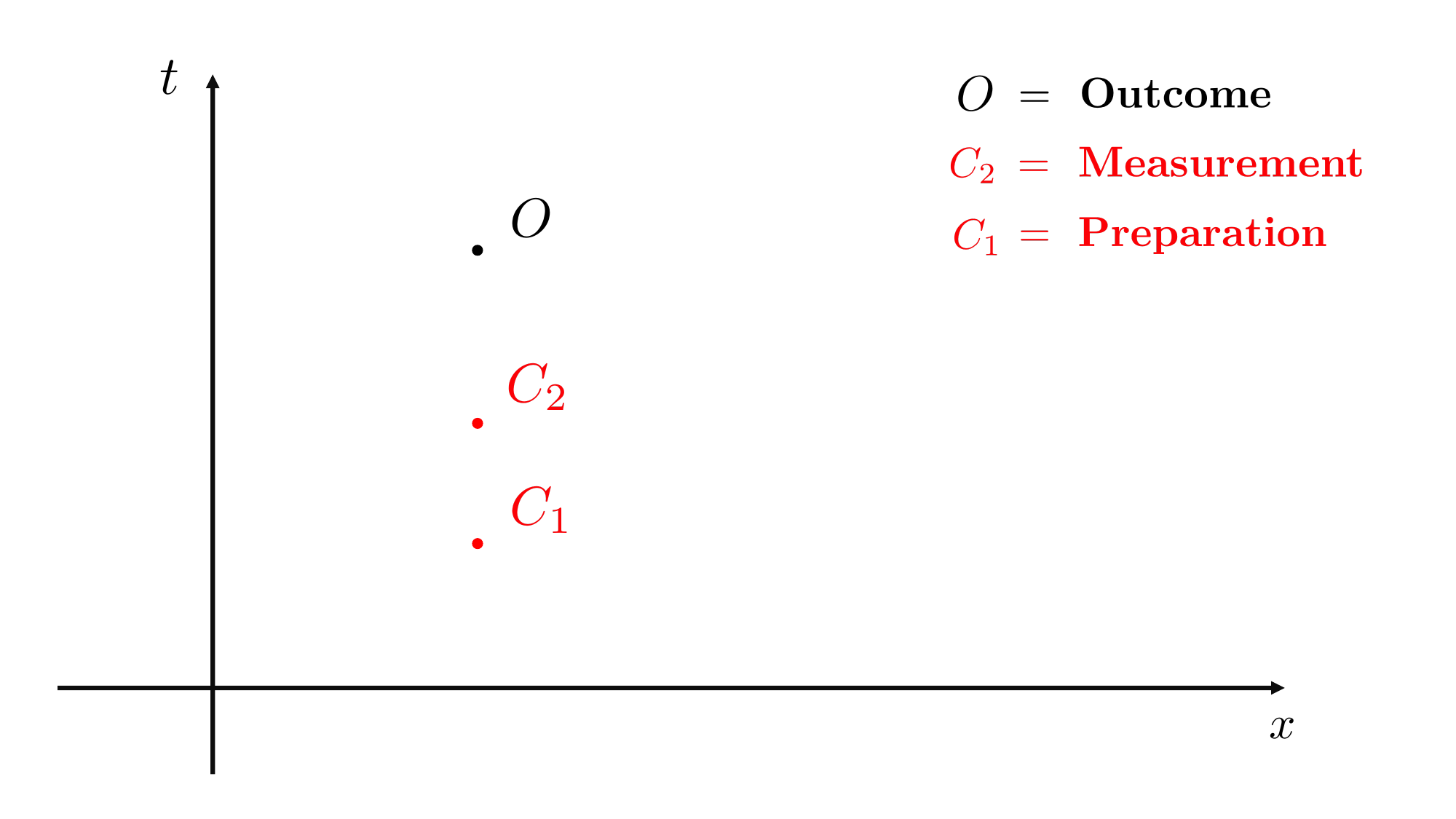}} 
	\caption{A prepare-and-measure experiment. The values taken by the controlled variables $C_1$ and $C_2$ specify the choices of preparation and measurement on the system, respectively. The value of the observed variable $O$ denotes the measurement outcome.}
	\label{ExperimentB}
\end{figure*}

Consider a prepare-and-measure experiment.  This has two controlled variables, $C_1$ and $C_2$, and one observed variable $O$ with the preorder relation $C_1 \prec C_2 \prec O$ (where $A \prec B$ means $A \preceq B$ and $B \npreceq A$).  The variable $C_1$ represents the choice of preparation for a physical system, $C_2$ represents a choice of measurement procedure, and $O$ represents the measurement outcome, as depicted in figure~\ref{ExperimentB}.

An \emph{ontological model} of a prepare-and-measure experiment $E$ is an ontic extension that satisfies the following additional assumptions.

\begin{enumerate}
\item \emph{$\lambda$-mediation}. The ontic states of the system mediate any correlation between the preparation and the measurement. More precisely, 
\begin{equation}
	q_E(O|\Lambda,C_1,C_2)=q_E(O|\Lambda,C_2).
\end{equation}
\item \emph{Measurement independence}: 
\begin{equation}
	q_E(\Lambda|C_1,C_2)=q_E(\Lambda|C_1).
\end{equation} 
This assumption is often motivated as an assumption of no-retrocausality\cite{LeiferPusey}.
\end{enumerate}

\begin{figure*}[]
\centering
{\includegraphics[scale=0.34]{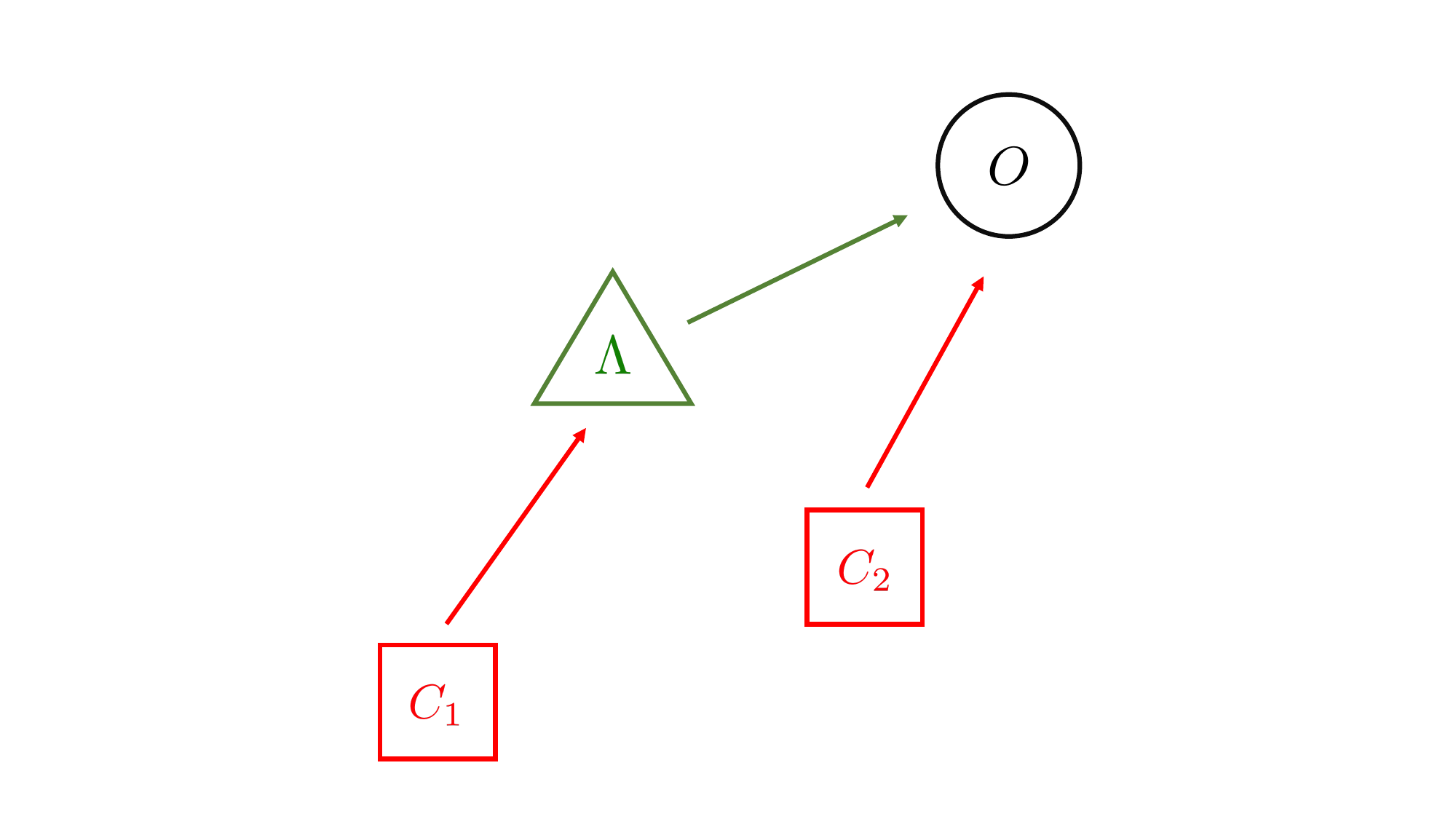}} 
\caption{Causal structure of the prepare and measure scenario in the standard ontological model framework. The directed acyclic graph above represents the causal relations between the operational and ontic variables. The extra assumptions of the standard ontological model framework \cite{Harrigan} involve restrictions on the possible causal structures associated to an experiment. These assumptions can be also obtained by imposing the causal Markov condition to the graph above \cite{Pearl}.}
\label{CausalStructure}
\end{figure*}
 
The additional assumptions imply that the experiment is described by the causal network illustrated in figure~\ref{CausalStructure}. We stress again that while these assumptions are needed to derive no-go theorems, we do not consider them to be part of our basic ontological framework, because they are not needed to define the notion of fine tunings. One main motivation for defining and characterizing operational fine tunings is to be equipped with the right framework, in future research, to study and develop new no-go theorems outside the ontological model framework. For example, considering an ontic extension that does not involve causal assumptions (as would be the case for an ontic extension representing a Block universe \cite{Minkowski1908}).

In order to provide a simple example of an ontological model, we can take classical Hamiltonian mechanics \cite{Goldstein}. In there the ontic state space is the phase space, whose points---specified by the positions and momenta of the particles---are the ontic states. 

\section{Examples of operational fine tunings}

\label{Examples}

In this section, we discuss three examples of no operational fine tuning assumptions, which have appeared previously in the literature.  This should help to motivate the general definition given in section \ref{Generalization}.

\begin{example}[Preparation noncontextuality \cite{Spekkens2005}\footnote{We focus on Spekkens' notion of preparation noncontextuality here for simplicity, but a similar account can be given of Spekkens' transformation and measurement noncontextuality.}]

Consider a prepare-and-measure experiment, as depicted in figure~\ref{ExperimentA}.  A preparation noncontextual ontic extension of an experiment is a model wherein, whenever two preparations are operationally equivalent, \textit{i.e.} whenever two preparations $c_1$ and $c'_1$ satisfy
\begin{equation}
	\label{OpEquiv}p(O|C_1 = c_1,C_2)=p(O|C_1 = c'_1,C_2) \;\;\;\;\;\;\;\;\;\;\;\;\; \forall C_2,O,
\end{equation}
the ontological descriptions are also the same
\begin{equation}
	\label{OnEquiv}q(O,\Lambda|C_1 = c_1,C_2)=q(O,\Lambda|C_1 = c'_1,C_2) \;\;\;\;\;\;\;\;\;\;\;\;\; \forall C_2,O.
\end{equation} 

Equation~\eqref{OnEquiv} may be an unfamiliar way of defining preparation noncontextuality, as the condition is usually written as
\[q(\Lambda|C_1 = c_1) = q(\Lambda|C_1 = c'_1).\]
However, when the ontic extension is an ontological model, the two definitions are equivalent\footnote{Up to issues with measure zero sets, which can be dealt with as in \cite{LeiferOntology}.}.  To see this, note that we can write
\[q(O,\Lambda|C_1,C_2) = q(O|\Lambda,C_1,C_2)q(\Lambda|C_1,C_2),\]
and then applying $\lambda$-mediation and measurement independence gives
\[q(O,\Lambda|C_1,C_2) = q(O|\Lambda,C_2)q(\Lambda|C_1).\]
Substituting these into equation~\eqref{OnEquiv} gives
\[q(O|\Lambda,C_2)q(\Lambda|C_1 = c_1) =  q(O|\Lambda,C_2)q(\Lambda|C_1 = c'_1),\]
and then dividing both sides by $q(O|\Lambda,C_2)$ gives the desired result.

The equivalence does not hold in a general ontic extension that is not an ontological model, and we view equation~\eqref{OnEquiv} as the appropriate generalization to this case.
 
An ontic extension that is not preparation noncontextual is called \emph{preparation contextual}. Preparation contextuality means that the operational equivalence of preparations is a result of a fine tuning of the ontic parameters that hides the difference between the preparation distributions of ontic states. 
Spekkens showed that any ontological model of quantum theory must be preparation contextual. 
\end{example}

\begin{example}[Parameter independence] 
The condition of local causality that is used to derive Bell's theorem \cite{Bell} can be decomposed into two assumptions: \emph{parameter independence} and \emph{outcome independence}

Consider the case of two spatially separated parties, Alice and Bob, who can each make one of two measurements on their systems (figure \ref{BellScenario}). We adopt the standard notation for a Bell scenario: Alice's and Bob's choices of measurements are denoted by the variables $X$ and $Y$ and their outcomes are $A$ and $B$ respectively. In the notation we have used so far, we would use $C_1$ and $C_2$ for $X$ and $Y$, and $O_1$ and $O_2$ for $A$ and $B$, but there is some value in following standard conventions. The no signaling condition says that Alice's measurement choice cannot influence Bob's outcome and vice versa, \textit{i.e.}
\begin{equation} \label{OpNoSignalling}
	\left\{\begin{array}{lr}
		p(A|X,Y=y)=p(A|X,Y=y') \\
		p(B|X=x,Y)=p(B|X=x',Y)
	\end{array}\right. 
\end{equation}
A natural explanation for this is that there is, in fact, no signaling at the ontological level.  This is a no fine tuning assumption as, were there signaling at the ontological level, it would take a special choice of parameters for the signaling to be inaccessible at the operational level.

\begin{figure*}[!htb]
	\centering
	{\includegraphics[scale=0.42]{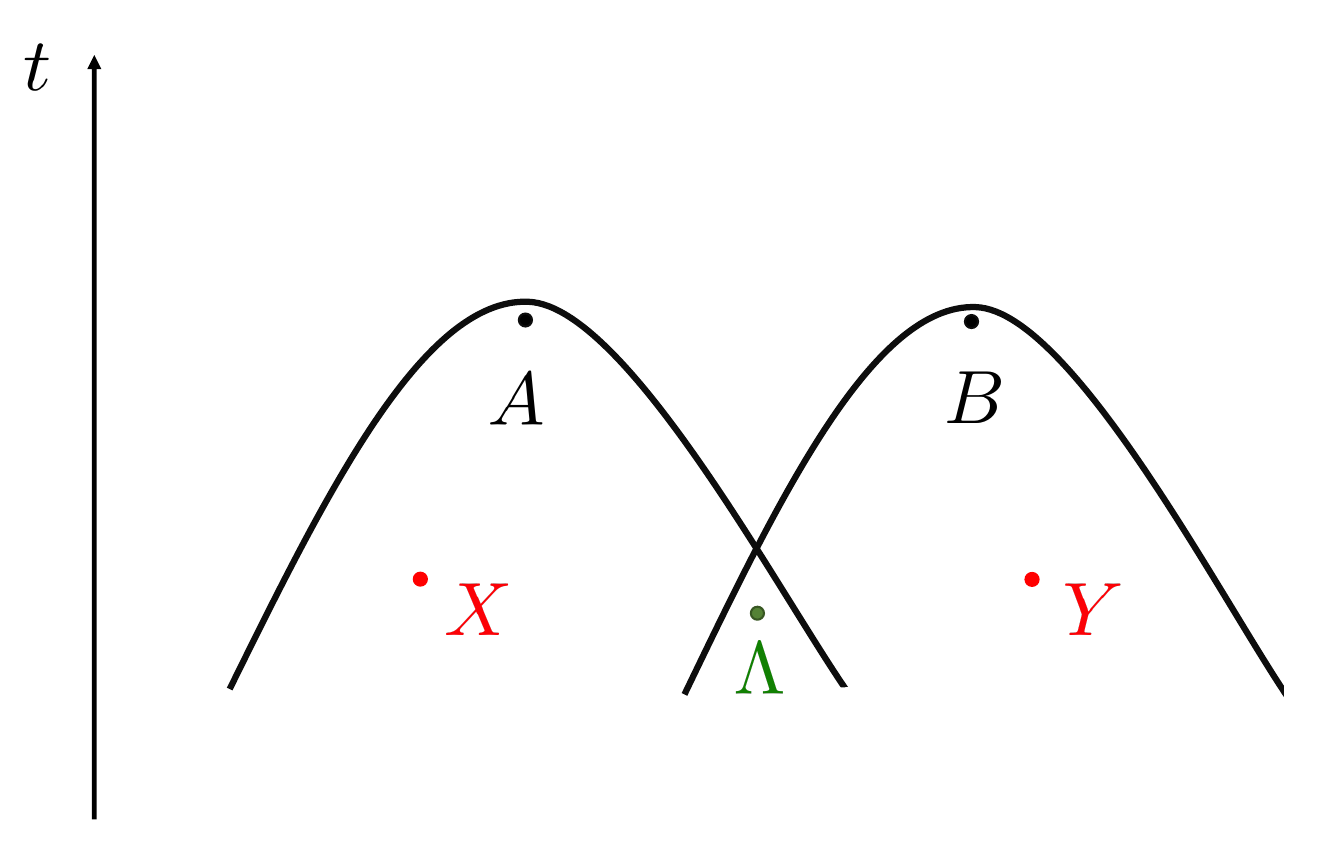}} 
	\caption{Bell Scenario. Alice and Bob perform two measurements $X,Y$ and obtain outcomes $A,B.$ The two light cones illustrate the space-like separation between Alice and Bob. Parameter independence is the requirement of no signaling between Alice's measurement(outcome) and Bob's outcome(measurement) at the ontological level.}
	\label{BellScenario}
\end{figure*}

The requirement that there is no signaling at the ontological level is parameter independence, which reads as follows.
  \begin{equation} \label{OnNoSignalling}
       \left\{\begin{array}{lr}
        q(A|\Lambda, X,Y=y)=q(A|\Lambda, X,Y=y') \\
        q(B|\Lambda, X=x,Y)=q(B|\Lambda, X=x',Y)
        \end{array}\right. ,
  \end{equation}
so we can view parameter independence as a no fine tuning assumption justified by the no-signaling principle.  
  
This is not enough to derive Bell's theorem, which requires the assumption of outcome independence as well.  Outcome independence reads as
\begin{equation}\label{OnOutcomeIndep}
      \left\{\begin{array}{lr}
      q(A|\Lambda, B=b,X,Y)=q(A|\Lambda,B=b', X,Y) \\
      q(B|\Lambda, A=a,X,Y)=q(B|\Lambda, A=a',X,Y).
      \end{array}\right.
\end{equation}
Together, these two assumptions imply local causality, which reads as
\begin{equation}
	q(A,B|\Lambda,X,Y) = q(A|\Lambda, X)q(B|\Lambda, Y).
\end{equation}
Bell's theorem states that the statistical predictions of quantum theory are inconsistent with the predictions of an ontological model satisfying local causality.
\end{example}

\begin{example}[Time symmetry] 
The time symmetry fine tuning is a more recent discovery \cite{LeiferPusey}, and will probably be less familiar.  We refer the reader to \cite{LeiferPusey} for a more careful discussion than is possible here.  An \emph{operationally} time symmetric theory is a theory in which, for every experiment $E$, there is another experiment $E'$ with a reversed time direction that yields the same probabilities.  The theories that describe our universe are \emph{not} operationally time symmetric because it is possible to send a signal into the future but not into the past.  However, we can restrict attention to the non-signaling sector of a theory, which means that we only consider the set of experiments that cannot be used to send a signal into the future.  The non-signaling sector of quantum theory is, in fact, operationally time symmetric.  The no fine tuning principle implies that the best explanation for this is that the ontological accounts of these experiments are time symmetric as well.

More precisely, we consider a preparation device that prepares a system and has both a controllable input and an observed output.  Let $C_1$ be the controlled input to the preparation and $O_1$ its observed output.  In addition, let $C_2$ be the input to a measurement device and $O_2$ its output. A theory has \emph{operational} time symmetry if for every experiment $E$ involving a choice of preparation $C_1=c_1$ with associated outcome $O_1=o_1$ and a choice of measurement $C_2=c_2$ with associated outcome $O_2=o_2$, there exists another experiment $E',$ where the choice of preparation is $C_1=c_2$ with associated outcome $O_1=o_2$ and the choice of measurement is $C_2=c_1$ with associated outcome $O_2=o_1,$ that provides the same statistics as $E,$ \textit{i.e.}
	\begin{align}
	\label{OpTimeSymmetry} 	
	&p_{E}(O_1 = o_1,O_2 = o_2|C_1 = c_1,C_2 = c_2) \notag\\ &= p_{E'}(O_1 = o_2,O_2 = o_1|C_1 = c_2,C_2 = c_1).
	\end{align}
This setup is depicted in figure~\ref{TimeSymmetryExperiment}. 

\begin{figure*}[!htb]
	\centering
	{\includegraphics[scale=0.42]{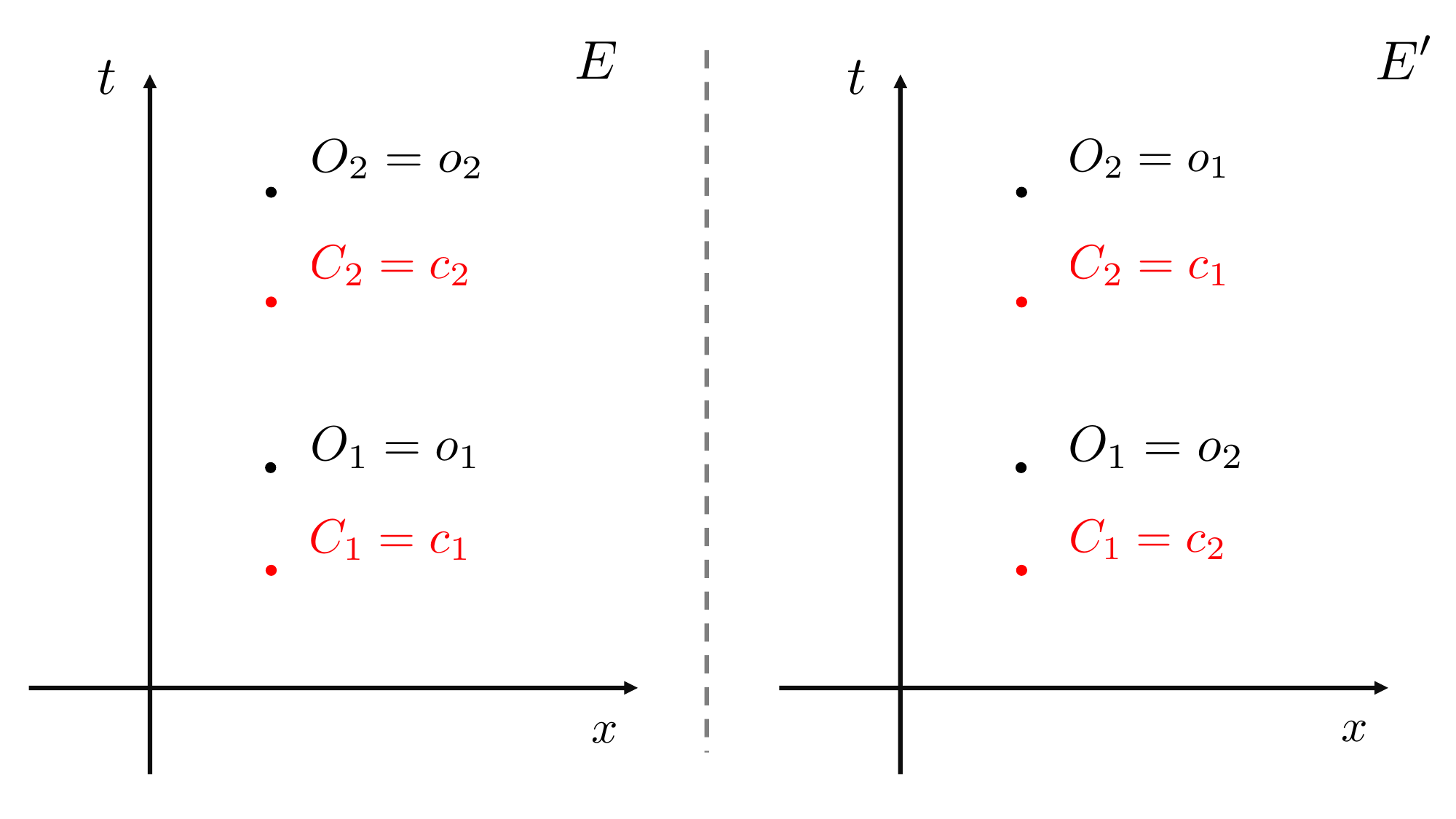}}
	\caption{Operational time symmetry. An operational theory is operationally time symmetric if for every experiment $E$ there exists another experiment $E'$, where the roles of the preparation and measurement are switched, that provides the same statistics as $E$.}
	\label{TimeSymmetryExperiment}
\end{figure*}

In this case, the no fine tuning assumption is that these two experiments should also satisfy \emph{ontological time symmetry}, which reads
	\begin{align} \label{OnTimeSymmetry} &q_{E}(O_1 = o_1,O_2 = o_2,\Lambda|C_1 = c_1,C_2 = c_2)\notag\\ &=q_{E'}(O_1 = o_2,O_2 = o_1,k(\Lambda)|C_1 = c_2,C_2 = c_1),
	\end{align} 
where $k$ is a one-to-one map $k:\mathfrak{L} \rightarrow \mathfrak{L}',$ with $\mathfrak{L},\mathfrak{L}'$ being the two ontic state spaces on which the ontic extensions of the two experiments $E$ and $E'$ are defined.  The map $k$ is included because, even in classical physics, time reversal transforms the ontic state.  For example, to time-reverse a trajectory in phase space requires inverting the momentum of each particle.

It is proven in \cite{LeiferPusey} that there is no ontological model of quantum theory that is ontologically time-symmetric for every experiment that has operational time-symmetry. 
\end{example}

\section{Classical Processings of Experiments}

\label{Proc}

In the examples just given, no fine tuning assumptions were identified by \textit{classically} processing the variables in an experiment in certain ways to identify operational equivalences. For example, we set controlled variables to different values, discarded observed variables, and interchanged the roles of preparations and measurements. Intuitively, a classical processing consists of actions that the experimenter can perform on the experiment, where ``classical'' means that the processings only act on the (classical) controlled and observed variables known to the experimenter. They cannot consist of, \textit{e.g.}, superchannels that act on quantum channels.  They cannot even involve changing the causal order between the controlled and corresponding observed variables in the experiment, \textit{e.g.} they cannot create signals backwards in time if such signals are not present in the original experiment.
In order to define operational fine tunings more generally, we first have to develop a general framework for describing these processings.  We start with processings that act on the experiments of the operational theory, and then describe how they are represented on the ontic extension.

\subsection{Operational processings}

Let us start by specifying that any experiment $E$ has a \emph{type}, which states the number of and cardinalities of the controlled and observed variables. Recall from section \ref{Operational} (see also figure~\ref{ExperimentA}) that the variables also form a specific pattern in space-time that, in principle, should also be a part of the specification of the type of $E$.  However, we will generally leave the specification of the pattern implicit and just focus on the probabilities assigned to the variables.

\begin{figure*}[!htb]
	\centering
	{\includegraphics[scale=0.465]{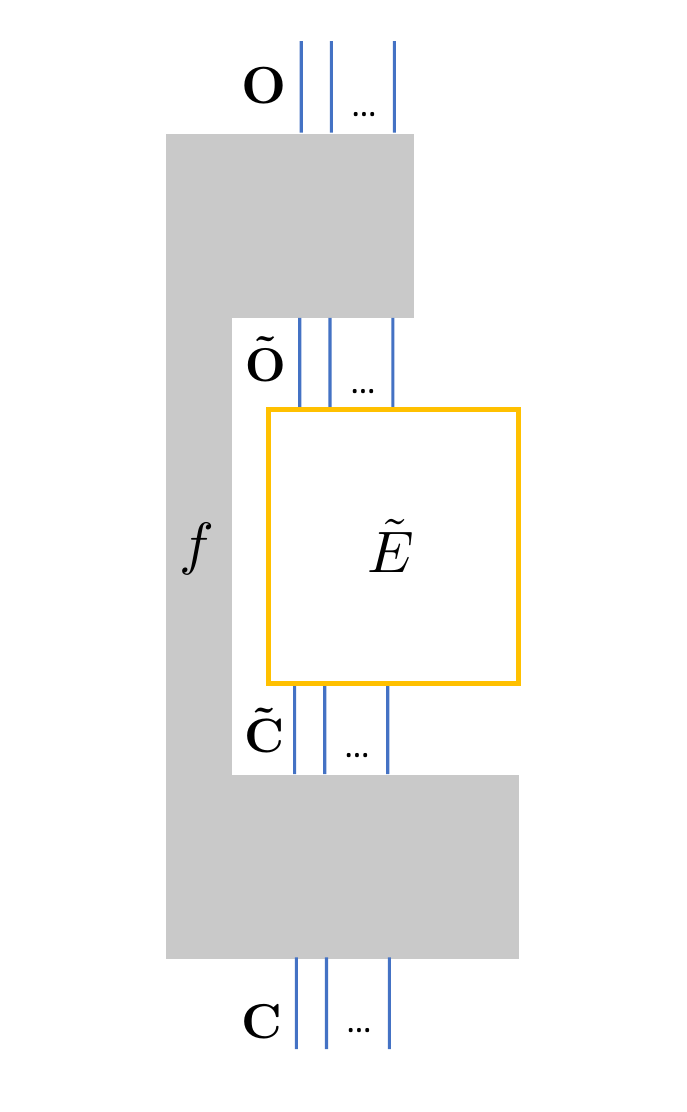}} 
	\caption{Diagrammatic representation of the classical processing $f$ of the experiment  $\tilde{E}.$ }
	\label{Diagram}
\end{figure*}

A classical processing acts on a specific type of experiment and produces a (possibly different) type of experiment. 
If a processing $f$ acts on an experiment $\tilde{E}$ to produce an experiment $E$ then we write $f(\tilde{E}) = E$.  In addition to a possible modification of the preorder associated with $\tilde{E}$,  the processing $f$ must specify a rule that produces the probability distribution of $E$ from the probability distribution of $\tilde{E}$.  Suppose that $\tilde{E}$ has controlled variables $\bm{\tilde{C}}$ and observed variables $\bm{\tilde{O}}$ with probability distribution $p_{\tilde{E}}(\bm{\tilde{O}}|\bm{\tilde{C}})$ and that $E$ has controlled variables $\bm{C}$ and observed variables $\bm{O}$ with probability distribution $p_{E}(\bm{O}|\bm{C})$.  To process $\tilde{E}$, we can perform a pre-processing, which takes the controls $\bm{\tilde{C}}$ of $\tilde{E}$ and produces the controls $\bm{C}$ for $E$.  This processing could be probabilistic, so in general it is specified by a conditional probability distribution $p_f(\bm{C}|\bm{\tilde{C}})$.  After observing $\bm{\tilde{O}}$, we can post-process those variables to produce $\bm{O}$.  Again, this can be probabilistic, so it would be specified by a conditional probability distribution $p_f(\bm{O}|\bm{\tilde{O}})$.  The action of the processing $f$ on the probability distribution $p_{\tilde{E}}(\bm{\tilde{O}}|\bm{\tilde{C}})$ -- represented in figure \ref{Diagram} -- is then given by

\begin{equation}
	\label{eq:ProcNC}
	p_{E}(\bm{O}|\bm{C}) = \sum_{\bm{\tilde{C}},\bm{\tilde{O}}} p_f(\bm{O}|\bm{\tilde{O}})p_{\tilde{E}}(\bm{\tilde{O}}|\bm{\tilde{C}}) p_f(\bm{\tilde{C}}|\bm{C}).
\end{equation}


\begin{figure}[]
\centering\offinterlineskip
\subfloat[][\label{ExampleProcessing}]
{\includegraphics[scale=0.23]{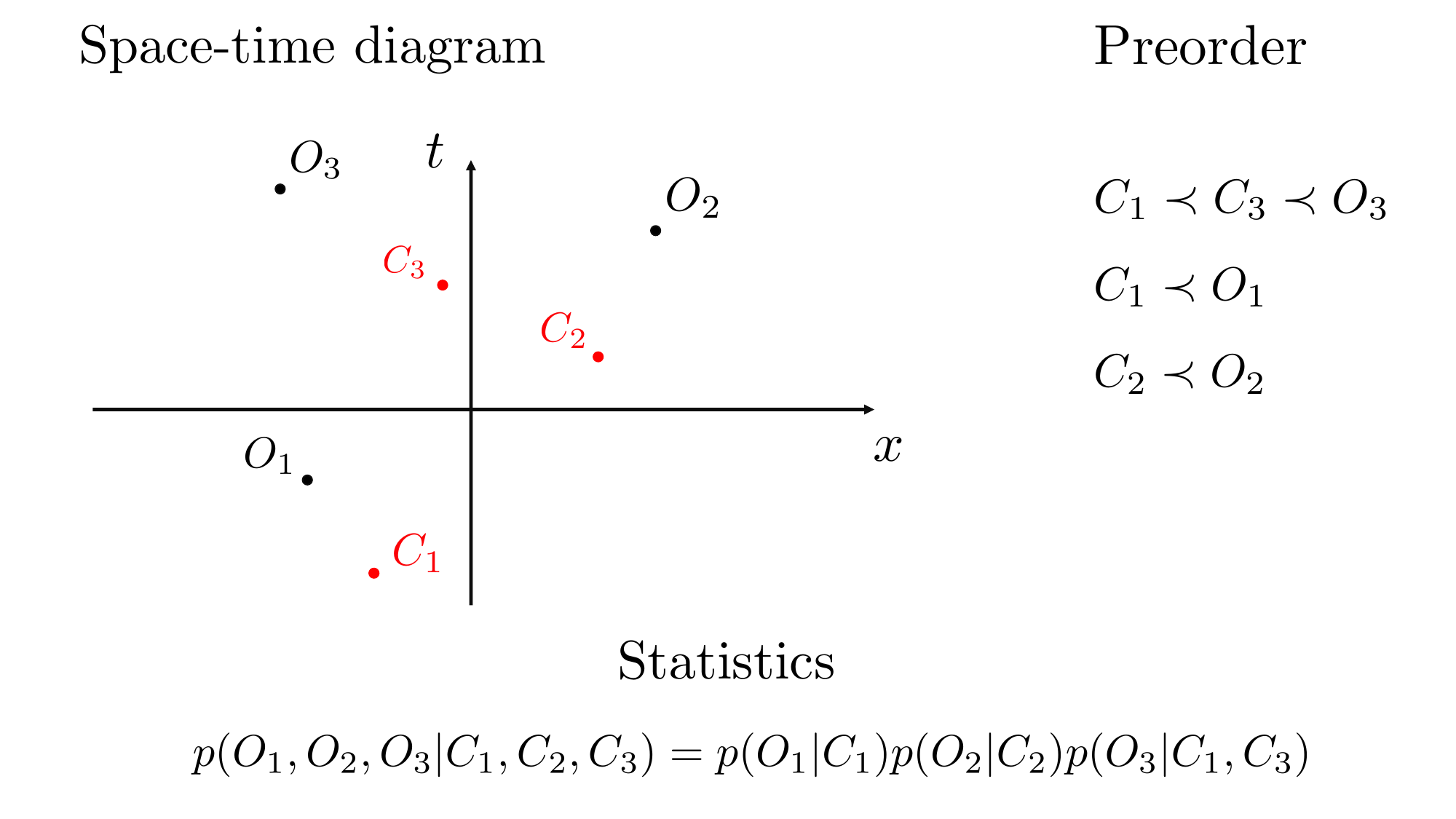}}
\centering
\subfloat[][\label{ExampleProcessingDeleting}]
{\includegraphics[scale=0.23]{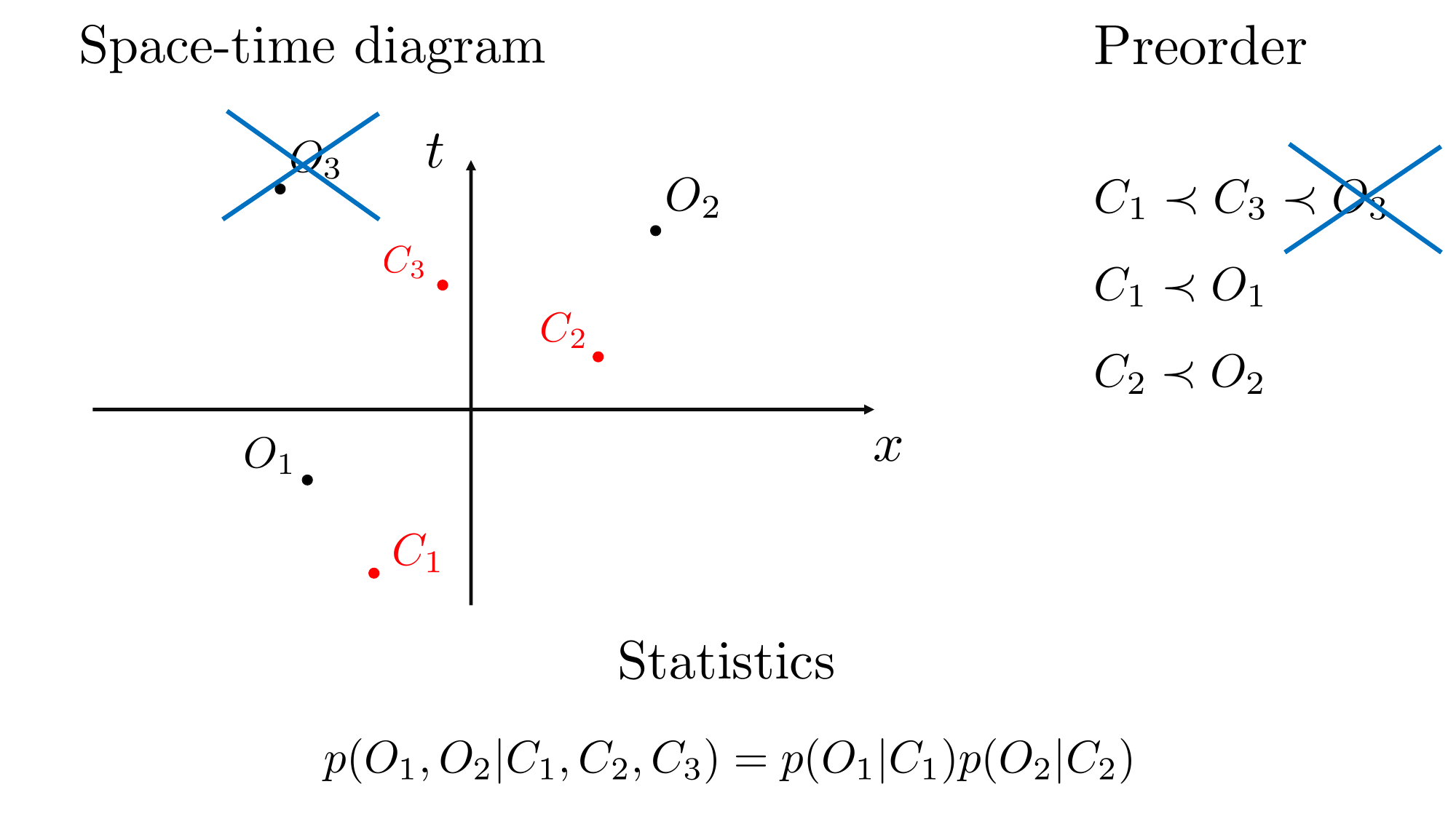}}\hfill 
\centering
\subfloat[][\label{ExampleProcessingAdding}]
{\includegraphics[scale=0.23]{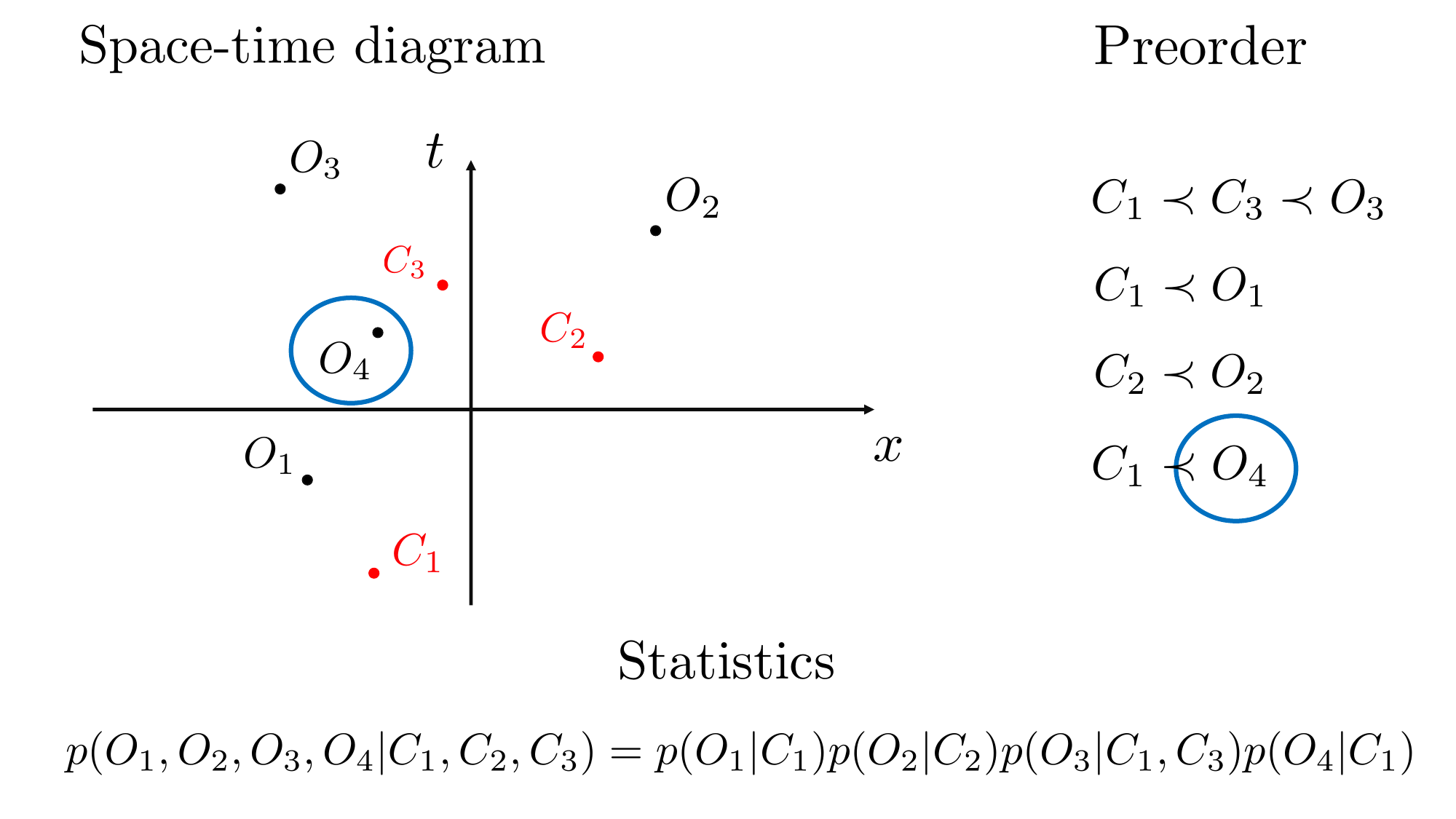}}
\centering
\subfloat[][\label{ExampleProcessingReplacing}]
{\includegraphics[scale=0.23]{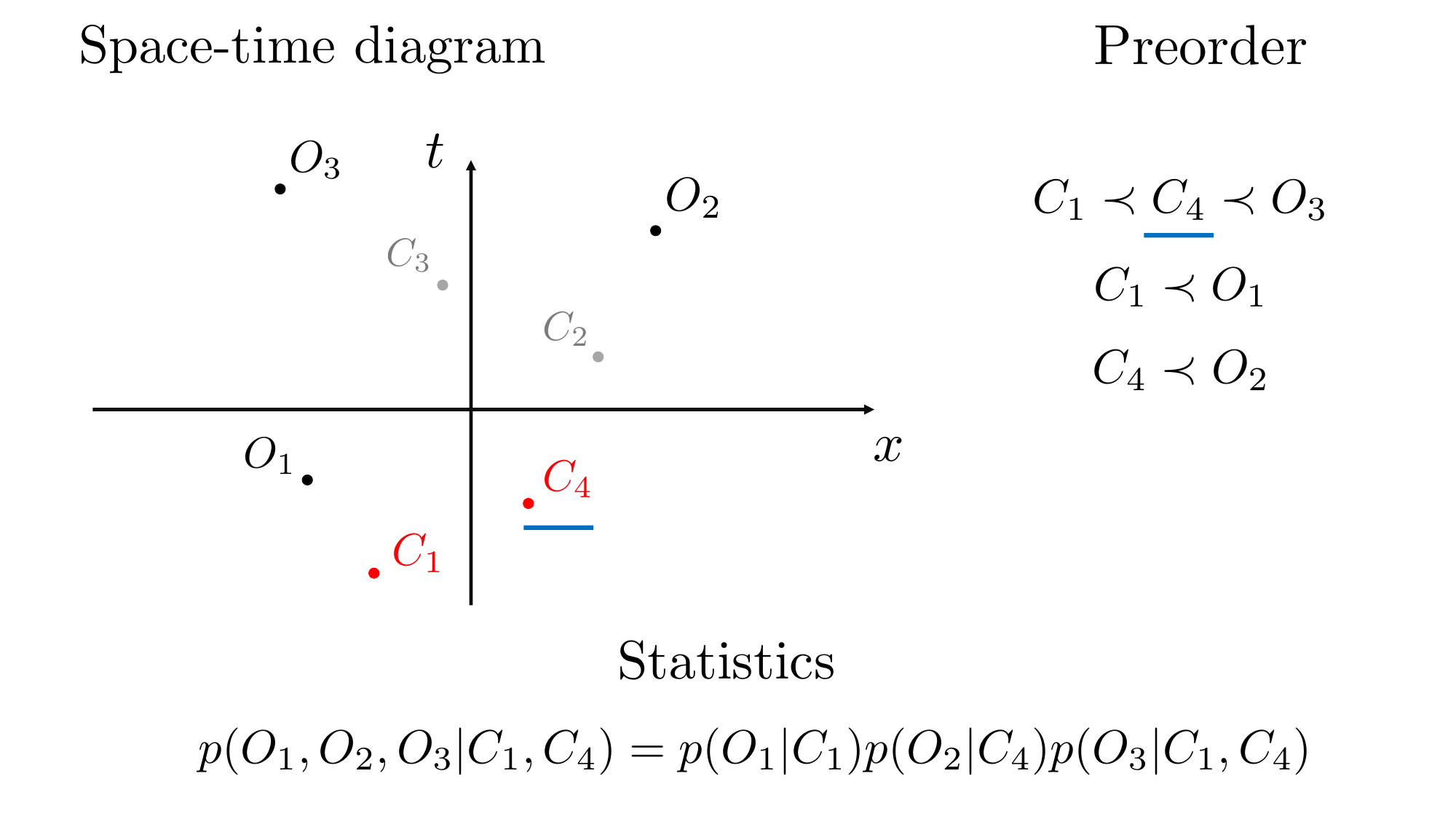}}
\caption{Examples showing the action of the primitive classical processings. Figure \ref{ExampleProcessing} represents the space-time diagram, preorder and conditional probabilities associated with a basic experiment. Figure \ref{ExampleProcessingDeleting}  represents the case where the processing consists of deleting the observed variable $O_3$, figure \ref{ExampleProcessingAdding} of adding the observed variable $O_4$, and figure \ref{ExampleProcessingReplacing} of replacing the controlled variables $C_2,C_3$ with $C_4$.}

\label{Examples_processings}
\end{figure}

In this work, we consider classical processings that are defined, in general, as a combination of three primitive processings (two concerning the post-processing and one the pre-processing), that we now describe. 
Each primitive processing consists of two ingredients: a modification of the conditional probabilities of observed variables given controlled variables and a modification of the preorder associated with the experiment.\footnote{ We recall that the preorder specifies the causal structure associated with the experiment. The specification of the new causal structure provides information on the causal dependencies between the operational variables and so on how the conditional probabilities of observed variables given controlled variables can be written -- see the examples in figure \ref{Examples_processings}.} 

\begin{enumerate}[label=(\alph*)]


\item \textit{Deleting an observed variable}. 
An experimenter can discard one of the observed variables.
Suppose the observed variables are $\bm{\tilde{O}}$ and the experimenter wants to remove $\tilde{O}_j$.  Let $\bm{\tilde{O}}_{-j}$ denote $\bm{\tilde{O}}$ with the variable $\tilde{O}_j$ removed.  The action of the discarding process $f=d_j$ in equation \eqref{eq:ProcNC} reads as follows, 
\[p_{d_j}(\bm{O} = \bm{o}|\bm{\tilde{O}} = \bm{\tilde{o}}) = \delta_{\bm{o},\bm{\tilde{o}}_{-j}},\]
where  $\bm{O}$ has the same number and cardinality of variables as $\bm{\tilde{O}}_{-j}$. 
The modification of the preorder here simply consists of the removal of the observed variable. Notice how this does not affect the causal relationship between the other variables. Figure \ref{ExampleProcessingDeleting} shows an example of an experiment subjected to the primitive processing associated with the removal of an observed variable and what this entails for the statistics $p_E({\bm{O}|\bm{C}})$ of the experiment.

This post-processing characterizes the case of parameter independence (equations \eqref{OpNoSignalling}), as we will show in the next section.

\item  \textit{Adding an observed variable.}
An experimenter can add an observed variable. 
Suppose the observed variables are $\bm{\tilde{O}}$ and the experimenter adds $\tilde{O}_j$.  Let $\bm{\tilde{O}}_{+j}$ denote $\bm{\tilde{O}}$ with the variable $\tilde{O}_j$ added.  The action of the adding process $f=a_j$ in equation \eqref{eq:ProcNC} reads as follows,  
\[p_{a_j}(\bm{O} = \bm{o}|\bm{\tilde{O}} = \bm{\tilde{o}}) = \delta_{\bm{o},\bm{\tilde{o}}_{+j}},\]
where  $\bm{O}$ has the same number and cardinality of variables as $\bm{\tilde{O}}_{+j}$. 
The modification of the preorder here simply consists of the addition of the observed variable, where the new variable can depend on any variable that is in its causal past. Notice how this does not induce new causal relationships between the other variables. If one deletes this new observed variable the preorder associated with the experiment returns to what it was.
Figure \ref{ExampleProcessingAdding} shows an example of an experiment subjected to the primitive processing associated with the addition of an observed variable and what this entails for the statistics $p_E({\bm{O}|\bm{C}})$ of the experiment.

We will see in the next section that by combining the removal and addition of observed variables it is possible to implement a post-processing that swaps the values of two observed variables, as in the case of operational time symmetry (equations \eqref{OpTimeSymmetry}). 




\item \textit{Replacing controlled variables}.
An experimenter can replace a subset of controlled variables with one controlled variable.

Suppose the controlled variables are $\bm{\tilde{C}}$ and the experimenter replaces a subset of them $\{\tilde{C}_j,\dots,\tilde{C}_{j+k}\}$ with the controlled variable $\tilde{C}_{jk}$, whose cardinality is not bigger than the sum of the cardinalities of the replaced variables.  Let $\bm{\tilde{C}}_{jk}$ denote $\bm{\tilde{C}}$ after the replacement.  The action of the replacing process $f=r_j$ in equation \eqref{eq:ProcNC} reads as follows,  
\[p_{r_j}(\bm{\tilde{C}} = \bm{\tilde{c}}|\bm{C} = \bm{c}) = \delta_{\bm{c},\bm{\tilde{c}}_{jk}},\]
where  $\bm{C}$ has the same number and cardinality of variables as $\bm{\tilde{C}}_{jk}$. 
The modification of the preorder here consists of placing the new controlled variable in the causal past of the replaced variables. Such variable preserves the causal relationships that the replaced variables had with the other variables. The simple case where only one variable is replaced corresponds just to move a controlled variable arbitrary far in the past. Figure \ref{ExampleProcessingReplacing} shows an example of an experiment subjected to the primitive processing associated with the replacement of a subset of controlled variables and what this entails for the statistics $p_E({\bm{O}|\bm{C}})$ of the experiment.

Let us stress that the replacement of controlled variables does not introduce new controls that are not already present in the experiment. It does not correspond for the experimenter to find out that there is a new knob in the experimental apparatus to use (it is a replacement, not an addition). It just corresponds to reduce the amount of control that the experimenter has. 

A special case of replacement is when the new variable $\tilde{C}_{jk}$ only takes one value (\textit{i.e.} it has cardinality $1$ --  it is a so called \textit{singleton}). Two important considerations about singletons are in order. First, we assume, as part of the primitive processings, that a singleton can always be deleted. Via the replacement processing, this can lead to the deletion of controlled variables (an experimenter can always choose to remove some controls). 
Second, notice that the case where the new singleton variable $\tilde{C}_{jk}$ replaces a single variable $\tilde{C}_j$ corresponds to the pre-processing that sets a controlled variable $\tilde{C}_j$ to a specific value $\tilde{c}$.  We denote this processing with $e_{j,c}$ and in equation \eqref{eq:ProcNC} it reads as follows,
\begin{equation}\label{ProcSet}p_{e_{j,c}}(\bm{\tilde{C}_j} = \bm{\tilde{c}}|\bm{C} = \bm{c}) = \delta_{\bm{c}_{-j},\bm{\tilde{c}}_{-j}}\delta_{c_j,\tilde{c}}.\end{equation}
This pre-processing characterizes the case of parameter independence (equations \eqref{OpNoSignalling}), as we will show in the next section.
We will also show that replacing controlled variables can be used to implement the pre-processing that swaps the values of two controlled variables, as in the case of operational time symmetry (equations \eqref{OpTimeSymmetry}).  


\end{enumerate}

Let us stress again that, given the primitive classical processings, any general processing is given by a combination of them.
Notice that, in the most trivial case, this includes doing nothing to the variables, \textit{i.e.} implementing an identity processing $\mathbb{I}$.  Formally, this is given by the pre-processing $p_{\mathbb{I}}(\bm{\tilde{C}}=\bm{\tilde{c}}|\bm{C}=\bm{c}) = \delta_{\bm{\tilde{c}},\bm{c}},$ and the post-processing $p_{\mathbb{I}}(\bm{O} = \bm{o}|\bm{\tilde{O}} = \bm{\tilde{o}}) = \delta_{\bm{o},\bm{\tilde{o}}}$.
Notice also that, in general (even if the examples we consider do not exhibit this), the pre and post-processings can be connected, \textit{i.e.} the post-processing be dependent on the pre-processing: what happens to the observed variables depends on what happened to the controlled variables.  In particular, this can be done as follows. One first introduces new observed variables $\bm{\tilde{O}}_D$ that depend on controlled variables $\bm{\tilde{C}}_D$, and then new observed variables $\bm{D}$ that depend on $\bm{\tilde{O}}_D$. At this point one deletes $\bm{\tilde{O}}_D$ and $\bm{\tilde{C}}_D$. This results in 
\[
	\label{eq:ProcGen}
	p_{E}(\bm{O}|\bm{C}) = \sum_{\bm{\tilde{C}},\bm{\tilde{O}},\bm{D}} p_f(\bm{O}|\bm{\tilde{O}},\bm{D})p_{\tilde{E}}(\bm{\tilde{O}}|\bm{\tilde{C}}) p_f(\bm{\tilde{C}},\bm{D}|\bm{C}).
\]

Processings have a compositional structure.  If the output type of processing $f$ is the same as the input type of processing $g$ then we can form the composition $g\circ f$, meaning ``first do $f$ and then do $g$''.  Suppose $f$ has pre-processing $p_f(\bm{\tilde{C}}|\bm{C})$ and post-processing $p_f(\bm{O}|\bm{\tilde{O}})$ and $g$ has pre-processing $p_g(\bm{C},|\bm{C}')$ and post-processing $p_g(\bm{O}'|\bm{O})$.  Then, the composite $g \circ f$ has pre-processing
\[p_{g \circ f}(\bm{\tilde{C}}|\bm{C}') = \sum_{\bm{C}} p_f(\bm{\tilde{C}}|\bm{C}) p_g(\bm{C}|\bm{C}'),\]

and post-processing

\[p_{g\circ f}(\bm{O}'|\bm{\tilde{O}}) = \sum_{\bm{O}} p_g(\bm{O}'|\bm{O})p_f(\bm{O}|\bm{\tilde{O}}).\]

Since processing is just classical manipulation of the operational variables, if we can implement the experiment $E$ then we can also implement the experiment $f(E)$. Thus, it is appropriate to demand that the theories under consideration are closed under processings.  This is formalized as follows.

Consider a set of experiments $\mathcal{E}$ and their associated operational probability distributions.  Let  $\mathcal{T}$ be the closure of $\mathcal{E}$ under processings, \textit{i.e.} if $E \in \mathcal{E}$ and $f$ is a processing then $f(E) \in \mathcal{T}$.  The set $\mathcal{T}$ together with the associated probability distributions is what we refer to as an \emph{operational theory}.  

Note that if $E \neq E'$ have the same operational probability distribution then we do not identify them as the same experiment.  In particular, if $E, E' \in \mathcal{E}$ and $E$ and $f(E')$ have the same operational probability distribution then they still are distinct experiments.  

The idea here is that experiments are not \emph{defined} by their probability distributions, but rather by a list of instructions for how to implement the experiment.  Those instructions can include classical processings of the variables in addition to the actual physical experiment, and those processings yield a new experiment.

\begin{definition}
An operational theory $\mathcal{T}$ is the closure of a set of experiments $\mathcal{E}$ under classical processings together with the associated probability distributions. 
\end{definition}

So far, we have only discussed processings that act on the operational variables.  We call these \emph{operational} processings to distinguish them from processings that act on the ontic extension.


\subsection{Ontological processings}

Suppose that an experiment $\tilde{E}$ with distribution $p_{\tilde{E}}(\bm{\tilde{O}}|\bm{\tilde{C}})$ is subjected to a processing $f$ to obtain the experiment $E$ with distribution $p_E(\bm{O}|\bm{C})$.  If $\tilde{E}$ has ontic extension $q_{\tilde{E}}(\bm{\tilde{O}},\Lambda|\bm{\tilde{C}})$ then the process $f$ must correspond to some process $h$ that operates at the ontological level, yielding an ontic extension $q_{E}(\bm{O},\Omega|\bm{C})$ of $E$.  Overloading the terminology, we call $h$ an \emph{ontic extension} of $f$.  
Note that, in general, we allow $\Lambda$ and $\Omega$ to live in different ontic state spaces $\mathfrak{L}$ and $\mathfrak{W}$.  For example, if $\tilde{E}$ is an experiment on a system composed of two subsystems, $A$ and $B$, then it might be the case that $\mathfrak{L}$ can be decomposed into a Cartesian product $\mathfrak{L} = \mathfrak{L}_A \times \mathfrak{L}_B$ of ontic states referring to system $A$ and those referring to system $B$.  Then, if $f$ discards system $B$, it would be appropriate for $h$ to discard $\Lambda_B$, yielding $\mathfrak{W} = \mathfrak{L}_A$. 

An ontic extension $h$ of $f$ must satisfy the following three properties:
\begin{enumerate}
	\item It must be a classical processing, acting at the ontological level, i.e. the action of $h$ is given via
	\[q_E(\bm{O},\Omega|\bm{C}) = \sum_{\bm{\tilde{C}},\bm{\tilde{O}},\Lambda} q_h(\bm{O},\Omega|\bm{\tilde{O}},\Lambda) q_{\tilde{E}}(\bm{\tilde{O}},\Lambda|\bm{\tilde{C}}) q_h(\bm{\tilde{C}}|\bm{C}),\]
	where $q_h(\bm{\tilde{C}}|\bm{C})$ and $q_h(\bm{O},\Omega|\bm{\tilde{O}},\Lambda)$ are conditional probability distributions.
	
	\item The processing $h$ must reproduce the operational processing $f$ when we marginalize over the ontic states, \textit{i.e.}
	\begin{align}
		q_h(\bm{\tilde{C}}|\bm{C}) & = p_f(\bm{\tilde{C}}|\bm{C}), \\
		\label{OnticExtension}
		\sum_{\Omega}q_{h}(\bm{O},\Omega|\Lambda,\bm{C}) & = p_{f}(\bm{O}|\bm{C}).
	\end{align}
	Note that equation~\eqref{OnticExtension} has the nontrivial implication that $q_h(\bm{O}|\Lambda,\bm{C}) = q_h(\bm{O}|\bm{C})$, \textit{i.e.} $\bm{O}$ is conditionally independent of $\Lambda$ given $\bm{C}.$ This is needed to ensure that $h$ reduces to $f$ at the operational level. 
	
	\item The ontic state $\Omega$ must be a \textit{sufficient statistic} for $\Lambda$ with respect to the observed variable $\bm{O}$. This means that 
	\begin{equation}
		\label{SS} q(\bm{O}|\Omega,\bm{C})=q(\bm{O}|\Lambda,\bm{C}).
	\end{equation}
	
	The idea behind this condition is that $\Lambda$ is supposed to explain the observed correlations between $\bm{\tilde{C}}$ and $\bm{\tilde{O}}$ in the experiment $\tilde{E}$.  Similarly, $\Omega$ is supposed to explain the observed correlations between $\bm{C}$ and $\bm{O}$ in $E$.  To count as the same ontic extension, $\Omega$ should explain the correlations between $\bm{C}$ and $\bm{O}$ in the same way that $\Lambda$ would have, and so any information about those correlations contained in $\Lambda$ should be preserved in $\Omega$.
	

	As an example, suppose that $E$ is an experiment on a composite system $AB$ and the ontic state space has a Cartesian product structure $\mathfrak{L}_{AB} = \mathfrak{L}_A \times \mathfrak{L}_B$, where $\mathfrak{L}_A$ represents properties of system $A$ and $\mathfrak{L}_B$ represents properties of system $B$.  If the ontic extension is doing a good job of explaining the observed probabilities then we expect that measurement outcomes on system $A$ are accounted for in terms of $\Lambda_A$ and those on system $B$ in terms of $\Lambda_B$.  The ontic state of $AB$ is $\Lambda = (\Lambda_A,\Lambda_B)$, so if we discard system $B$ as part of the processing $f$, we would expect to be able to discard $\Lambda_B$ as part of the action of the ontic extension $h$. However, we would not expect to be able to discard $\Lambda_A$, as it is supposed to be doing the job of explaining the observed probabilities for system $A$. The sufficient statistics condition ensures that if $\Lambda_A$ is indeed correlated with the operational variables for system $A$ then $\Omega$ must also be correlated in the same way, \textit{e.g.} $\Omega=\Lambda_A$. In this case the sufficient statistics condition would read as $q(A|\Omega,C)=q(A|\Lambda,C),$ where $A$ is the observed variable associated with the outcomes of the properties of system $A$ and $C$ is some arbitrary controlled input variable of the experiment.
	
	In general, our experiment needs not to have a neat subsystem structure, but we still want to ensure that $\Omega$ is playing the same explanatory role as $\Lambda$.  Thus, if $\Lambda$ is correlated with the operational variables of $E$ then $\Omega$ should be correlated with them in the same way.
	Without the sufficient statistics condition, an ontic extension $h$ of an operational processing $f$ could always just delete the ontic state $\Lambda$ entirely. 

\end{enumerate}


\section{General operational fine tunings} 

\label{Generalization}

In this section, we give our general definition of the operational no fine tuning condition and show that the three examples given in section \ref{Examples} are instances of it.  The general idea is that if there are two experiments that satisfy an operational equivalence, then that equivalence should be preserved at the ontological level.  Let us define what we mean by operational equivalence. 

Suppose that there exist two experiments $E$ and $E'$, with distributions $p_E(\bm{\tilde{O}}|\bm{\tilde{C}})$ and $p_{E'}(\bm{\tilde{O}}'|\bm{\tilde{C}}')$ in our operational theory, and two classical processings $f$ and $f'$, such that\footnote{Notice that, in this section, we call $E$ and $E'$ rather than $\tilde{E}$ and $\tilde{E}'$ the experiments before the processings in order to soften the notation, considering the numerous formulas that will be involved.} 


\begin{flalign} \label{OpFT}
	&\sum_{\bm{\tilde{C}},\bm{\tilde{O}}} p_{f}(\bm{O}|\bm{\tilde{O}})p_E(\bm{\tilde{O}}|\bm{\tilde{C}})p_{f}(\bm{\tilde{C}}|\bm{C}) \notag\\ & = \sum_{\bm{\tilde{C}}',\bm{\tilde{O}}'} p_{f'}(\bm{O}'|\bm{\tilde{O}}')p_{E'}(\bm{\tilde{O}}'|\bm{\tilde{C}}')p_{f'}(\bm{\tilde{C}}'|\bm{C}').  
\end{flalign}
We call such an equation an \emph{operational equivalence}.  The main idea of our no fine tuning condition is that there should be an analogous equation that holds in the ontic extensions of these experiments.

Note that operational equivalences come in various strengths.  In the parameter independence example, the no-signaling condition holds for \emph{all} experiments, and in the time-symmetry example \emph{every} experiment in the no-signaling sector of quantum theory has an operational time-reverse.  In contrast, the operational equivalences used in preparation noncontextuality only hold for given pairs of experiments.  Arguably, stronger operational equivalences, \textit{i.e.} those that hold for all experiments, are more likely to be universal physical principles, so violations of the no fine tuning condition based on them are more problematic. 


In order to define our no fine tuning condition, two additional assumptions about the structure of ontic extensions are needed. These form the condition of structure preservation, that reads as follows.

\emph{Structure preservation}.  Operational processings and ontological processings have structure.  In particular, there is an identity operational processing $\mathbb{I}$ and an ontological identity processing $\mathbb{I}_{\Lambda}$ that do nothing to any of the variables.  The identity operational processing, already introduced in the previous section, is specified by 
	\begin{align*}
		p_{\mathbb{I}}(\bm{\tilde{C}}|\bm{C}) & =\delta_{\bm{\tilde{c}},\bm{c}}, &
		p_{\mathbb{I}}(\bm{O}|\bm{\tilde{O}}) & =\delta_{\bm{o},\bm{\tilde{o}}}, 
	\end{align*}
	and the identity ontological processing is specified by
	\begin{align*}
		q_{\mathbb{I}_{\Lambda}}(\bm{\tilde{C}}|\bm{C}) & =\delta_{\bm{\tilde{c}},\bm{c}}, &
		q_{\mathbb{I}_{\Lambda}}(\bm{O},\Omega|\bm{\tilde{O}},\Lambda) & =\delta_{\bm{o},\bm{\tilde{o}}} \delta_{\omega,\lambda}. 
	\end{align*}	
	\begin{itemize}
		\item If the operational processing is $f = \mathbb{I}$ then its ontic extension is $h = \mathbb{I}_{\Lambda}$.
		\item If the operational processing $f''$ is obtained by composing $f$ with $f'$, which we denote $f'' = f' \circ f$, then the corresponding ontic extensions $h$, $h'$ and $h''$, should satisfy $h'' = h' \circ_{\Lambda} h$, where $\circ_{\Lambda}$ denotes composition of ontological processings.
	\end{itemize}  
	
	The intuition behind these conditions is fairly straightforward.  If we do absolutely nothing to a physical system then we would expect that nothing happens to its ontological description.  If we compose two processings then the ontic description should be transformed by applying the two processings one after the other.

We are now in a position to state our no fine tuning condition.

\begin{figure*}[htb!]
\centering
{\includegraphics[scale=0.4]{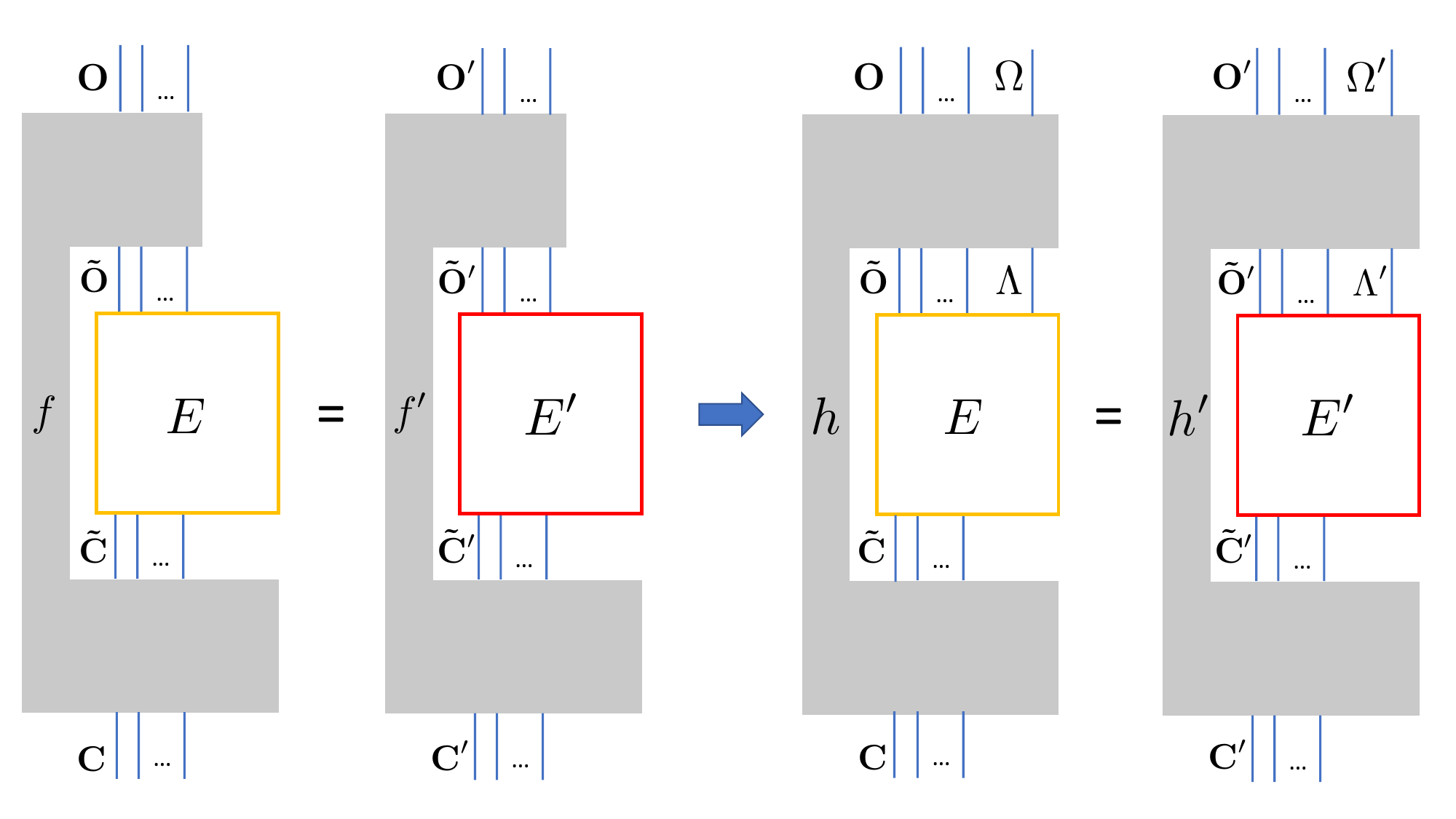}} 
\caption{A schematic representation of the no fine tuning condition. The condition demands that an operational equivalence, possibly involving classical processings (represented in the figure by the grey boxes) on two experiments $E,E',$ implies a consistent (\textit{i.e.} satisfying the properties (i), (ii), (iii) and the condition of structure preservation) ontological equivalence.  The above representation is meant to be read in terms of the conditional probabilities associated with the experiments, \textit{i.e.} as in equations \eqref{OpFT} and \eqref{OnFT}.}
\label{SchemeFT}
\end{figure*}

\begin{definition}{\emph{No fine tuning condition}.}
\label{NoFT}
	Let $\mathcal{T}$ be an operational theory and suppose that it has an ontic extension $\mathcal{T}_{\Lambda}$ that satisfies the condition of structure preservation. Suppose that $\mathcal{T}$ has an operational equivalence between experiments $E$ and $E',$ for some operational processings $f$ and $f'$, as in equation \eqref{OpFT}.
	
	The ontic extension $\mathcal{T}_{\Lambda}$ is \emph{not fine tuned} with respect to this operational equivalence if
	\begin{flalign}
	\label{OnFT}
		&\sum_{\bm{\tilde{C}},\bm{\tilde{O}},\Lambda} q_{h}(\bm{O},\Omega|\bm{\tilde{O}},\Lambda)q_E(\bm{\tilde{O}},\Lambda|\bm{\tilde{C}})p_{h}(\bm{\tilde{C}}|\bm{C}) \notag\\ &= \sum_{\bm{\tilde{C}'},\bm{\tilde{O}'},\Lambda'} q_{h'}(\bm{O}',\Omega'|\bm{\tilde{O}}',\Lambda')q_{E'}(\bm{\tilde{O}}',\Lambda'|\bm{\tilde{C}}')p_{h'}(\bm{\tilde{C}}'|\bm{C}'), 
	\end{flalign}
	where $q_E$, $q_{E'}$, $h$ and $h'$ are the ontic extensions of $p_E$, $p_{E'}$, $f$ and $f'$ respectively.
	
	Otherwise the ontic extension is \emph{fine tuned} with respect to the operational equivalence.	
\end{definition}
Figure~\ref{SchemeFT} gives a schematic representation of the no fine tuning condition.

\vspace{5mm}

\newpage
Although the no fine tuning condition might seem rather complicated, it captures the intuition that we started with.  The most natural explanation for an operational equivalence is that it is an actual equivalence at the ontological level.  If not, then the parameters of the ontic extension have to be chosen in a special way, \textit{i.e.} they are fine tuned.

Notice that the no fine tuning condition is always satisfied by the trivial ontic extension, in which the ontic and operational variables are uncorrelated, \textit{i.e.} \[q_E(\bm{O},\Lambda|\bm{C})=p_E(\bm{O}|\bm{C})q(\Lambda),\] 
and in which the ontic extension of a processing $f$ does nothing to $\Lambda$ and acts the same way as $f$ on the operational variables.  Thus, there is no question of proving fine tuning for an ontic extension without additional assumptions.

If we want our ontological descriptions to explain the observed correlations between the operational variables then the notion of an ontic extension is too weak, since, as in the trivial ontic extension, it allows the ontic states to be completely uncorrelated from the operational variables.  Thus, additional assumptions are warranted. The additional causal assumptions of the ontological models framework are needed to rule out trivial ontic extensions and prove no-go theorems. 

In the simple case of prepare and measure experiments, trivial ontic extensions are ruled out by the $\lambda-$mediation assumption.  For more general experiments, we leave the question of the most appropriate additional assumptions open for future research.  For the present purpose of defining operational fine tunings and explaining the assumptions of existing no-go theorems in terms of them, such additional assumptions are not needed.

\begin{figure}[]

\centering
\subfloat[][\label{a}]
{\includegraphics[scale=0.334]{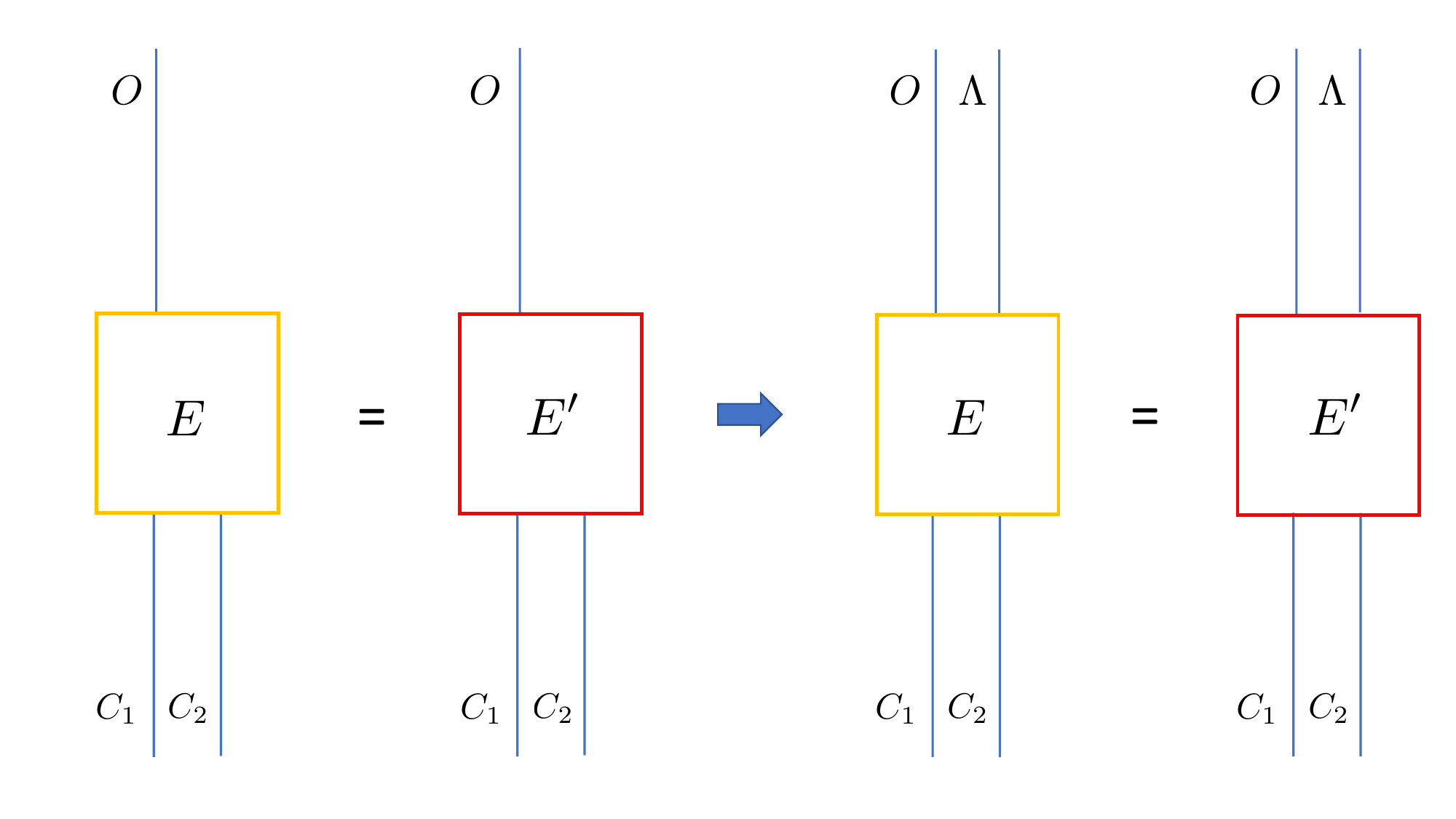}}\hfil 
\subfloat[][\label{b}]
{\includegraphics[scale=0.334]{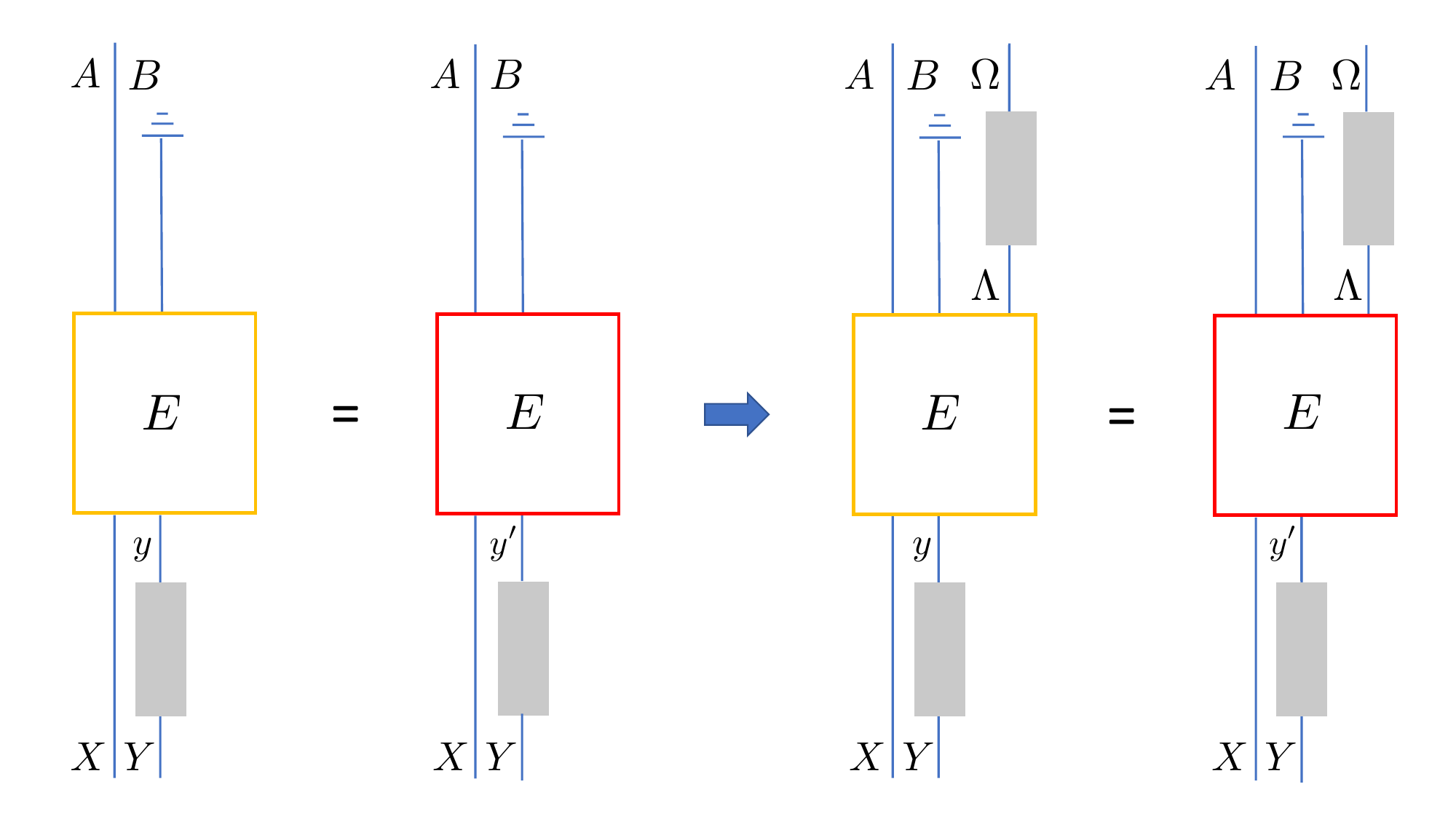}}\hfil 
\subfloat[][\label{c}]
{\includegraphics[scale=0.334]{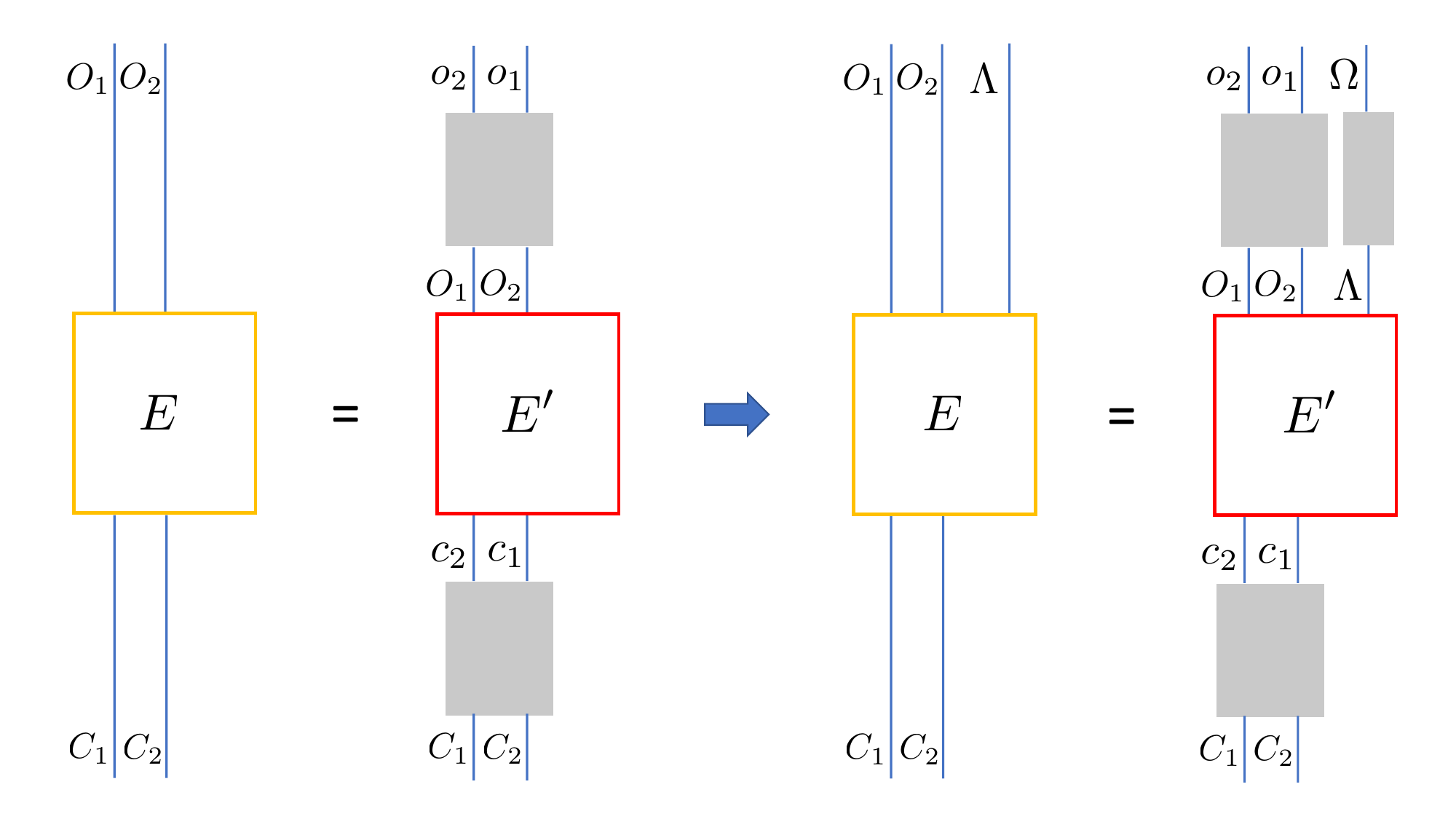}}

\caption{Three examples of no operational fine tuning requirements in the generalized framework. Figure \ref{a} represents the requirement of (preparation) noncontextuality, 
 figure \ref{b} of parameter independence (the first equation in \eqref{OpNoSignalling} and \eqref{OnNoSignalling}) and figure \ref{c} of time symmetry (here it is intended that $C_1=c_1$, $C_2=c_2$, $O_1=o_1$ and $O_2=o_2$ before the processings). The grey boxes represent the non-trivial classical processings that characterize each case.}

\label{Examples_figure}
\end{figure}

Let us now show how the equations \eqref{OpFT} and \eqref{OnFT} reduce to the known ones for the examples considered in the previous section.
Figure \ref{Examples_figure} schematically represents them in light of the new framework.

\begin{itemize}



\item \emph{Preparation noncontextuality}.  
In this case the classical processing is trivial, meaning that the pre and post processings correspond to the identity processing, \textit{i.e.} $p_{f}(O|\tilde{O})=\delta_{o,\tilde{o}},$ and $p_{f}(\tilde{C}|C)=\delta_{\tilde{c},c}$. The same happens with $f'$. Equation \eqref{OpFT} reduces to the equation \eqref{OpEquiv}, where the variable $C$ specifies which preparation is considered (more precisely, it is $C_1$ in equation \eqref{OpEquiv}, given that we also have $C_2$ indicating the measurement procedure). The different choices of preparations are specified by the labels of the experiments $E$ and $E'.$
By imposing the no fine tuning condition \ref{NoFT} we obtain \[\sum_{\Lambda} q_{h}(\Omega | \Lambda)q_E(O,\Lambda |C)=\sum_{\Lambda} q_{h'}(\Omega | \Lambda)q_{E'}(O,\Lambda |C).\] Given the structure preserving condition (ii), the post processing including ontological variables must be the identity. This means that $q_{h}(\Omega | \Lambda)=\delta_{\omega,\lambda} \;\; \forall i\in\{0,1\}$ and that $\Omega\equiv \Lambda.$ The same happens with $h'$. Therefore equation \eqref{OnFT} reduces to equation \eqref{OnEquiv} and the requirement of noncontextuality is obtained.

\item \emph{Parameter independence}.  
Let us first rewrite the left hand side of equation \eqref{OpFT} identifying $C_1=X, C_2=Y, O_1=A, O_2=B.$, \[  \sum_{\tilde{O}_1,\tilde{O}_2,\tilde{C}_1,\tilde{C}_2} p_{f}(A,B|\tilde{O}_1,\tilde{O}_2)p_E(\tilde{O}_1,\tilde{O}_2|\tilde{C}_1,\tilde{C}_2)p_{f}(\tilde{C}_1,\tilde{C}_2|X,Y).\]  The right hand side of equation  \eqref{OpFT} reads the same apart from $f'$ instead of $f.$
The pre-processing that sets the value of the variable $Y$ to $y$ and leaves $X$ the same is the one that we denoted with $e_{j,c}$ in equation \eqref{ProcSet}, which identifies the values of $\tilde{C}_1$ with the values of $X$ and sets $Y$ to $y$.   
The post-processing that corresponds to marginalize out $B$ is the elementary process (a) that consists of deleting the observed variable $B$.  Combining these pre and post processings in the left hand side of equation \eqref{OpFT} leads to $p(A|X,Y=y).$ The same happens for the right hand side of equation \eqref{OpFT}, with the only exception being the controlled variable $Y$ getting the value $y'$ instead of $y$, thus yielding to $p(A|X,Y=y').$  In conclusion we obtain the first equation in \eqref{OpNoSignalling}.
Notice how, in this case, the classical processings do not entail any non-trivial modification of the preorder. 
The no fine tuning condition \ref{NoFT} leads to \begin{flalign*}\sum_{\Lambda,\tilde{O}_1} q_{h}(A,\Omega | \Lambda,\tilde{O}_1)q(\tilde{O}_1,\Lambda |X,Y=y)= \sum_{\Lambda,\tilde{O}_1} q_{h}(A,\Omega | \Lambda,\tilde{O}_1)q(\tilde{O}_1,\Lambda |X,Y=y').\end
{flalign*} The classical law of total probability then yields to \begin{equation}\label{passage}q(A,\Omega |X,Y=y)=q(A,\Omega |X,Y=y').\end{equation} Notice that if we sum over $A$ on both sides we obtain that $q(\Omega|X,Y=y)=q(\Omega|X,Y=y').$ By the analogous procedure on the second equation in \eqref{OpNoSignalling} we obtain $q(\Omega|X=x,Y)=q(\Omega|X=x',Y).$ Together these two equations mean that $q(\Omega|X,Y)=q(\Omega).$ 
This condition is useful for our purposes because we can divide both sides of equation \eqref{passage} by it and obtain \[q(A |\Omega, X,Y=y)=q(A|\Omega,X,Y=y').\] The condition of parameter independence in equation \eqref{OnNoSignalling} then follows from the condition (iii) of sufficient statistics.

\item \emph{Time symmetry}. 
In this case $f$ is the identity processing and so for obtaining the left hand side of equation \eqref{OpTimeSymmetry} we use the same treatment that we used for the case of preparation noncontextuality. For the right hand side we need to permute the values of the two controlled variables and the values of the two observed variables.
The post-processing that swaps the values of the observed variables is obtained, in terms of  the primitive processings (a) and (b), as follows. One first adds an observed variable $O_1$ that takes value $\tilde{o}_2$, and then deletes $\tilde{O}_2$. Similarly, one then adds an observed variable $O_2$ that takes value $\tilde{o}_1$, and then deletes $\tilde{O}_1$. 
The pre-processing that swaps the values of the controlled variables is obtained, in terms of the primitive processing (c), as follows. One first replaces the controlled variable $\tilde{C}_1$ with one variable $C_1$ that takes value $\tilde{c}_2$. Similarly, one then replaces the controlled variable $\tilde{C}_2$ with one variable $C_2$ that takes value $\tilde{c}_1$. 
Notice how in this case the modification of the preorder is not trivial -- one moves the observed variable $O_1$ far in the future and the controlled variable $C_2$ far in the past -- and how this allows to perform the swapping of the variables (we represent the modifications of the preorder and conditional probabilities in figure \ref{CausalStructureTimeSymmetry}). 
With respect to equation \eqref{eq:ProcNC} we have
 \[\sum_{\tilde{C},\tilde{O}} p_{f'}(O_1,O_2|\tilde{O}_1,\tilde{O}_2)p_{E'}(\tilde{O}_1,\tilde{O}_2|\tilde{C}_1,\tilde{C}_2)p_{f'}(\tilde{C}_1,\tilde{C}_2|C_1,C_2),\] where $p_{f'}(O_1,O_2|\tilde{O}_1,\tilde{O}_2)=\delta_{o_1,\tilde{o}_2} \delta_{o_2,\tilde{o}_1}$ and $p_{f'}(\tilde{C}_1,\tilde{C}_2|C_1,C_2)=\delta_{\tilde{c}_1,c_2} \delta_{\tilde{c}_2,c_1},$ which means that in E', $C_1=c_2$, $C_2=c_1$ and $O_1=o_2$, $O_2=o_1$, thus yielding equation \eqref{OpTimeSymmetry}.
By imposing the no fine tuning condition \ref{NoFT} we obtain \begin{align*}& \sum_{\tilde{C},\tilde{O},\Lambda} q_{h}(O_1,O_2,\Omega|\Lambda,\tilde{O}_1,\tilde{O}_2)q_{E}(\tilde{O}_1,\tilde{O}_2,\Lambda|\tilde{C}_1,\tilde{C}_2)p_{h}(\tilde{C}_1,\tilde{C}_2|C_1,C_2) = \\ & \sum_{\tilde{C},\tilde{O},\Lambda} q_{h'}(O_1,O_2,\Omega|\Lambda,\tilde{O}_1,\tilde{O}_2)q_{E'}(\tilde{O}_1,\tilde{O}_2,\Lambda|\tilde{C}_1,\tilde{C}_2)p_{h'}(\tilde{C}_1,\tilde{C}_2|C_1,C_2).\end{align*} The left hand side can be treated in the same way as we treated the left hand side of the preparation noncontextuality case. For the right hand side we need to use the structure preserving condition (ii) for the composition of processings. First, we need to notice that  $f= f'\circ f'$, where we recall that $f$ is the identity processing and $f'$ is the processing that permutes inputs and outputs (so applied two times in sequence it gives back the original conditional probability). As a consequence of the structure preserving condition (ii) we then have that $h' \circ h'$ will also be equal to the identity processing, and that the identity processing will correspond to h, as h is the ontic processing associated to f. This means that $h'$ is an involution like $f'$, thus guaranteeing that the ontological post-processing reads as \begin{flalign*}q_{h'}(O_1,O_2,\Omega|\Lambda,\tilde{O}_1,\tilde{O}_2)= 
 \delta_{\omega,k(\lambda)}\delta_{o_1,\tilde{o}_2}\delta_{o_2,\tilde{o}_1},\end{flalign*}
where $k$ is a one to one function, $k:\mathfrak{L} \rightarrow \mathfrak{L}',$ with $\mathfrak{L},\mathfrak{L}'$ being the two ontic state spaces on which $h$ and $h'$ are defined, and that the pre-processing reads as  $p_{h'}(\tilde{C}_1,\tilde{C}_2|C_1,C_2)=\delta_{\tilde{c}_1,c_2} \delta_{\tilde{c}_2,c_1}.$ 
The condition of time symmetry in equation \eqref{OnTimeSymmetry} is thus recovered.

\begin{figure*}[!htb]
	\centering
	\subfloat[][\label{aa}]
{\includegraphics[scale=0.33]{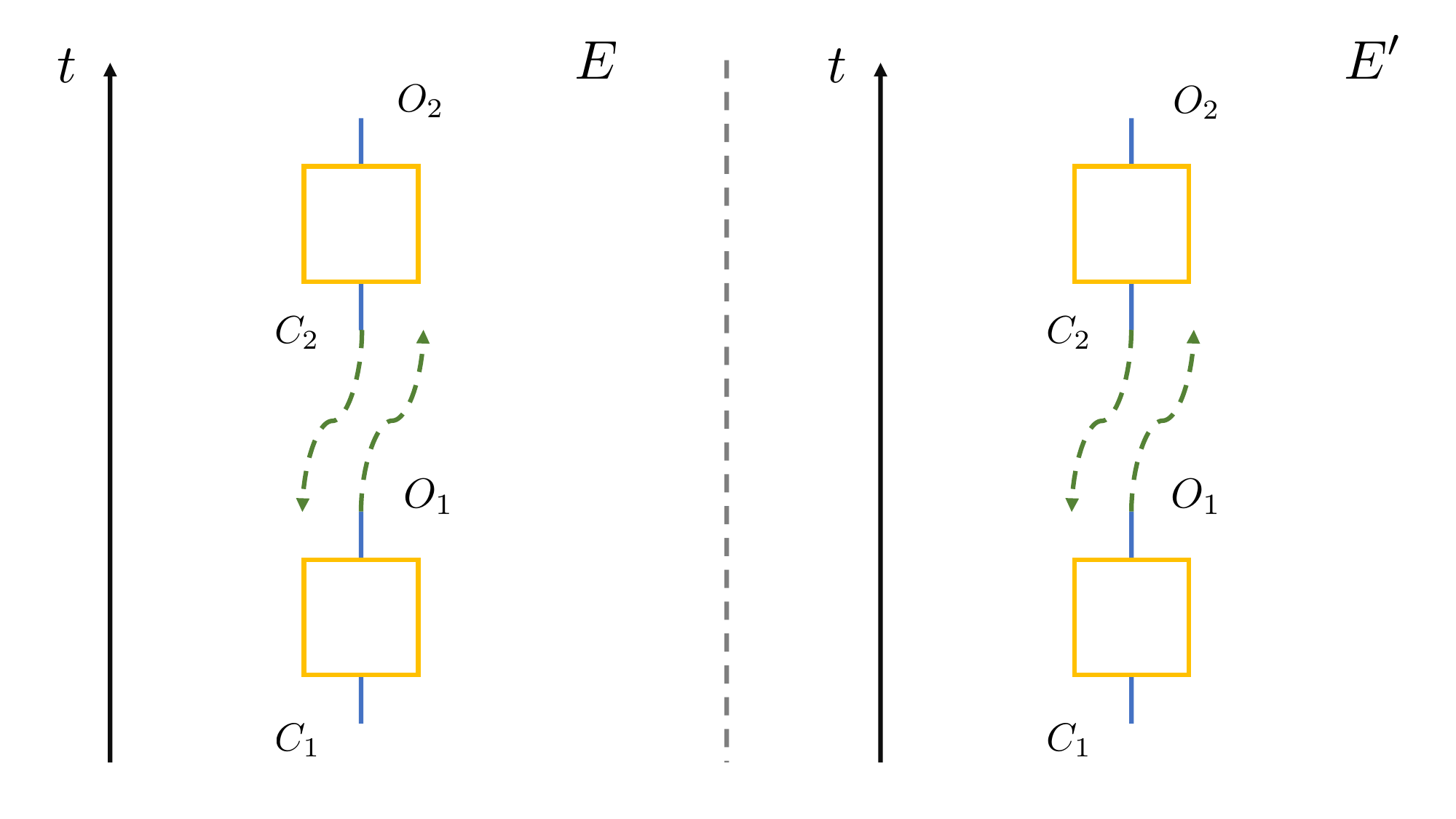}}\hfil 
\subfloat[][\label{bb}]
{\includegraphics[scale=0.33]{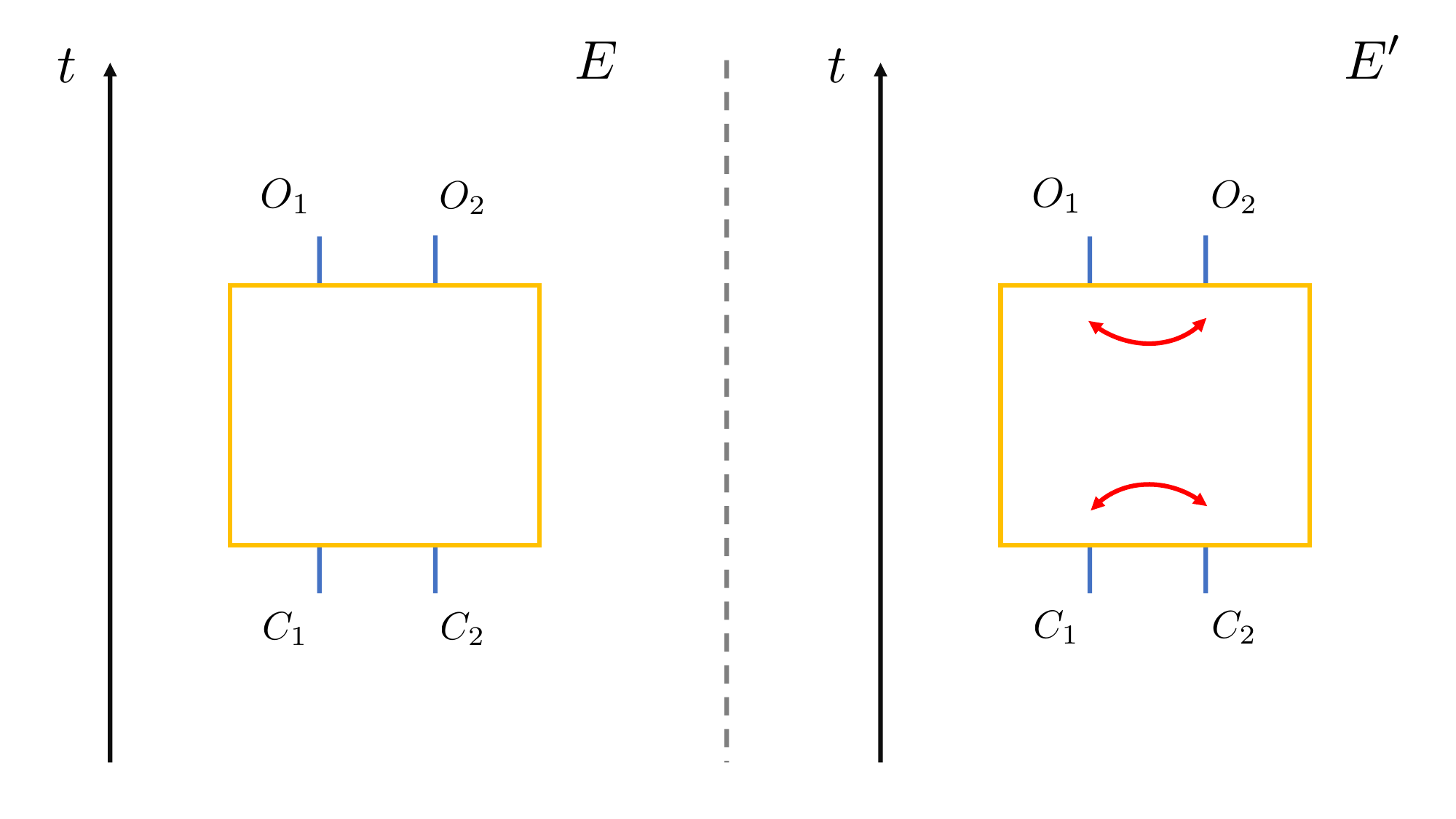}}
	\caption{Classical processing in the operational time symmetry example. Here we represent the modification of the preorder (figure \ref{aa}) and the modification of the conditional probabilities -- the swapping between the preparation variables and between the measurement variables (figure \ref{bb}).}
	\label{CausalStructureTimeSymmetry}
\end{figure*}

\end{itemize}


\section{The categorical framework}

\label{Cats}

The previous construction for generalizing operational fine tunings finds a natural formulation in the framework of category theory \cite{Leinster2014}. Let us briefly refresh the basics of category theory.
A category $\mathcal{C}=(\mathcal{O},\mathcal{M})$ is defined as a set $\mathcal{O}$ of objects and a set $\mathcal{M}$ of morphisms acting on them. Given two objects $a,b\in\mathcal{C},$ a morphism between them is denoted with $a\rightarrow b.$ By definition, in a category there is a notion of composition of morphisms -- a binary operation $\circ$ -- that satisfies the properties of associativity, \emph{i.e.} if $f:a\rightarrow b, \; g: b\rightarrow c$ and $h: c\rightarrow d,$ then $h\circ (g\circ f)= (h\circ g) \circ f,$ and the existence of the identity morphism for every object $x$, \emph{i.e.} $\mathbb{I}_x : x\rightarrow x$ such that for every morphism $f: a\rightarrow b$ we have $\mathbb{I}_b \circ f = f = f \circ \mathbb{I}_a.$ A map between categories that preserves their structure is called a functor. More precisely, a functor $\mathcal{G}$ between two categories $\mathcal{C}_1$ and $\mathcal{C}_2$ associates an object $x\in \mathcal{C}_1$ to an object $\mathcal{G}(x)\in \mathcal{C}_2$ and associates to each morphism $f:x\rightarrow y\in \mathcal{C}_1$ a morphism $\mathcal{G}(f): \mathcal{G}(x)\rightarrow \mathcal{G}(y)$ that preserves the identity morphism, $\mathcal{G}(\mathbb{I}_x)=\mathbb{I}_{\mathcal{G}(x)} \;\; \forall \; x\in \mathcal{C}_1,$ and the composition rule, $\mathcal{G}(g\circ f) = \mathcal{G}(g)\circ \mathcal{G}(f) \;\; \forall \; f: x\rightarrow y, \; g: y\rightarrow z \in \mathcal{C}_1.$ 
Notice the similarity between the properties of a functor and the structure preserving conditions of the previous section.

Let us now define the operational and ontological categories.
The former refers to all the possible experimental statistics associated to experiments (without any operational constraint), while the latter refers to the corresponding ontological representations.

\begin{definition}{\emph{Operational Category}.}
\label{OpCat}
The operational category $\mathcal{O}p=(\mathcal{P}, \mathcal{F})$ is a category where the objects are sets of conditional probability distributions of observed variables given controlled variables, $\{p(\bm{O}|\bm{C})\}\in \mathcal{P}.$ Each set includes conditional probability distributions with the same number of observed variables and controlled variables, and those variables have the same cardinality. The morphisms in the category are classical processings between those sets of probabilities, \[f:\{p(\bm{\tilde{O}}|\bm{\tilde{C}}) \} \rightarrow \{ p(\bm{O}| \bm{C})\}.\] We denote the set of such morphisms as $\mathcal{F}.$ 
\end{definition}

The motivation for defining an object in the operational category as a set of conditional probability distributions with the same number of observed variables $\bm{O}$ and controlled variables $\bm{C}$, with the same cardinality, is that, operationally, conditional probability distributions $p(\bm{O}|\bm{C})$ within the same set provide equal amount of information about the nature of the system subjected to the experiment.  As a clarifying example, let us imagine, in quantum theory, an experiment $E_1$ involving two preparations of a qubit with a two-outcome measurement and an experiment $E_2$ on a qutrit, again involving two preparations and a two-outcome measurement. According to definition \ref{OpCat}, the probability distributions $p_{E_1}(O|C)$ and $p_{E_2}(O|C)$ belong to the same object in the category. This may seem counterintuitive, because these probabilities refer to experiments involving different kinds of systems. However, the crucial point is that, from an operational perspective, these two probabilities do not allow to infer such distinction. In particular, we would need extra variables and/or greater cardinality to realize that $E_2$ involves a qutrit.  


Notice, from definition \ref{OpCat}, that a precise conditional probability distribution $p(\bm{O}|\bm{C})$ associated to a particular experimental setup can be defined in the categorical language as a morphism from the trivial object of the category -- which is the conditional probability distribution with one element $O$ and one element $C,$ $p(O|C),$ both with cardinality one -- to the desired conditional probability distribution $p(\bm{O}|\bm{C}).$ These morphisms are called \emph{states}.
An operational theory $\mathcal{T}$ is associated to a subcategory $\mathcal{O}p_{\mathcal{T}}$ of the operational category $\mathcal{O}p,$ where only a set of objects $\mathcal{P}_{\mathcal{T}}\subset\mathcal{P}$ and morphisms $\mathcal{F}_{\mathcal{T}}\subset\mathcal{F}$ that map between them are considered. For example, quantum theory is associated to the subcategory of the operational category with the constraint that objects are composed by sets of conditional probabilities consistent with the statistics predicted by quantum theory, where the controlled and observed variables refer to events in experimental scenarios that are possible in quantum theory -- \emph{e.g.} turning a knob to select the orientation of a Stern-Gerlach magnet and detect where, on the screen, the particle outputs.

\begin{definition}{\emph{Ontological Category}.}
\label{OnCat}
The ontological category $\mathcal{O}n=(\mathcal{P}_{\Lambda}, \mathcal{H})$ is a category where the objects are sets of conditional probability distributions of observed and ontic variables given controlled variables, $\{q(\bm{O}, \Lambda|\bm{C})\}\in \mathcal{P}_{\Lambda}.$ Each set includes conditional probability distributions with the same number of observed variables, ontic variables and controlled variables, and those variables have the same cardinality. The morphisms in the category are classical processings between those probabilities, \begin{flalign*}h:\{q(\bold{\tilde{O}},  \Lambda|\bold{\tilde{C}})\} \rightarrow \{q(\bold{O}, \Omega|\bold{C})\}.\end{flalign*} These morphisms must satisfy the sufficient statistics condition as in the property (iii) of section \ref{Proc}, meaning that \begin{flalign*}q( \bold{O}|  \Omega, \bold{C}) = q( \bold{O}|  \Lambda,  \bold{C})\end{flalign*} for every $q( \bold{O}|  \Omega, \bold{C})\in \{q( \bold{O}|  \Omega, \bold{C})\}$ and for every object $\{q( \bold{O}|  \Omega, \bold{C})\}.$ We denote the set of such morphisms as $\mathcal{H}.$ 
\end{definition}


Given the definitions above,  the notion of ontic extension $\mathcal{T}_{\Lambda}$ of an operational theory $\mathcal{T}$ is associated to the existence of a functor from the operational subcategory $\mathcal{O}p_{\mathcal{T}}$ to the ontological category $\mathcal{O}n$. Therefore we can define the no fine tuning condition as follows.

\begin{definition}{\emph{No fine tuning condition in categorical terms}.}
\label{NoFTCat}
Let $\mathcal{O}p_{\mathcal{T}}$ be the operational subcategory associated with the operational theory $\mathcal{T}$ and suppose that  $\mathcal{G}:\mathcal{O}p_{\mathcal{T}}\rightarrow \mathcal{O}n$ is the functor associated with the ontic extension $\mathcal{T}_{\Lambda}$ of the theory. Suppose that in $\mathcal{O}p_{\mathcal{T}}$  an operational equivalence -- an equation between different states associated with experiments $E$ and $E',$ for some morphisms $f$ and $f'$ -- holds. The functor is \emph{not fine tuned} with respect to the operational equivalences if it preserves the equation -- now between states subjected to morphisms $h$ and $h'$ -- in the ontological category $\mathcal{O}n$. 
\end{definition}

The reason why it is natural to use the categorical framework for defining operational fine tunings is that the functors, by definition, preserve the structure between the operational and ontological categories and allow to easily include the necessary properties that we listed in the previous section.

\section{Discussion and conclusion}

\label{Conclusion}

In the framework of ontological models the nonclassical aspect of quantum theory that emerges from the popular no-go theorems is distilled in the notion of fine tuning. We have here provided a general mathematical framework for characterizing such notion. More precisely, we have rigorously defined operational fine tunings. 
An ontic extension of an operational theory is fine tuned with respect to an operational equivalence predicted by the theory if it does not preserve such equivalence. In other words, the parameters of the ontic extension have to be chosen in a special way to account for such operational equivalence, \textit{i.e.} they are fine tuned.
In the language of category theory, a functor -- associated with an ontic extension of the operational theory -- is fine tuned with respect to an operational equivalence predicted by the operational theory -- defined as an equation between different states subjected to possible morphisms in the corresponding operational category --  if the functor does not preserve the equation in the ontological category. 

Our framework both accommodates already known operational fine tunings -- arising from the violation of generalized noncontextuality, parameter independence and time symmetry -- and provides the mathematical ground for more general ones, in the sense that it accounts for fine tunings involving any generic classical processing on the experiments under consideration. We leave the study of novel operational fine tunings for future work. However, we already argue that an interesting class of possibly new fine tunings is the class formed by group symmetry fine tunings. Let us look, in particular, at the Lorentz symmetry group. When considering Bell's scenario, Lorentz invariance has already been shown to break at the ontological level for particular ontological models, like the ones of Bohmian mechanics \cite{BohmianMechanics} and collapse theories \cite{CollapseTheories}. It would be interesting to study whether this must be the case for any ontological model. 

We define fine tunings with respect to ontic extensions, that represent a more general ontological framework than the standard ontological model framework used to prove the popular no-go theorems \cite{Harrigan}. As a result, a value of our work is to categorize the assumptions in such no-go theorems in terms of the ones about realism (ontic extension), the ones about causality (\textit{e.g.} $\lambda-$mediation and measurement independence) and the ones about the requirement of no operational fine tuning (\textit{e.g.} noncontextuality or parameter independence).

An application and insight entailed by our framework regards the relationship between nonlocality and contextuality. It is usually claimed that the former is an instance of the latter. Within the Kochen-Specker contextuality literature \cite{KochenSpecker}, when considering the standard Bell scenario, the different sets of Alice's and Bob's measurements are treated as the different contexts of measurements and Bell inequalities become just an example of noncontextuality inequalities, as well expressed in the graph theoretic approach \cite{CSW,Acin2015}. Another framework where nonlocality is an instance of Kochen-Specker contextuality is provided by the sheaf theoretic approach \cite{Abramsky2011}. Regarding the generalized notion of contextuality\footnote{We recall that the assumption of Kochen-Specker noncontextuality is equivalent to the assumptions of measurement noncontextuality and outcome determinism for projective measurements. In quantum theory, the assumption of outcome determinism for projective measurements can be derived from preparation noncontextuality \cite{Spekkens2005}.} \cite{Spekkens2005}, nonlocality is seen to be an instance of preparation contextuality by looking at the standard Bell scenario in quantum theory, where any proof of nonlocality can be mapped to a proof of preparation contextuality by considering the prepare and measure scenario where the preparation is specified by the state Alice (Bob) steers Bob’s (Alice's) system to, and the measurement is Bob's (Alice's) measurement. 

In contrast to these arguments, our approach, that relies on the framework of ontic extensions rather than ontological models,  shows a neat difference between the locality (Bell's local causality) and the (generalized) noncontextuality assumptions. As already discussed, locality can be decomposed into the assumptions of parameter independence (equation \eqref{OnNoSignalling}) and outcome independence (equation \eqref{OnOutcomeIndep}).  In this work we showed that parameter independence can be justified by a requirement of no operational fine tuning. In this sense it represents the ``noncontextuality part'' of the assumption of locality. However, the additional assumption of outcome independence, which is crucially needed for having locality, has a purely causal nature. It cannot be justified by a requirement of no operational fine tuning. Therefore nonlocality can result from breaking either the  noncontextuality-like assumption of parameter independence or the purely causal assumption of outcome independence. On the contrary, contextuality involves a purely operational fine tuning, as shown by our framework. 

In terms of the statement that nonlocality is an instance of preparation contextuality, it is worth noticing that when, in Bell scenario, nonlocality corresponds to breaking the purely causal assumption of outcome independence, this coincides, in the prepare and measure scenario, to violate the assumption of $\lambda-$mediation -- one of the causal assumptions in the ontological model framework -- and \emph{not} the actual assumption of noncontextuality. In summary, one can say that nonlocality is an instance of preparation contextuality in the standard ontological model framework, if one means that a proof of the impossibility of a local ontological model implies a proof of the impossibility of a preparation noncontextual ontological model; however, the former does not imply a proof of the impossibility of a noncontextual ontic extension, because the impossibility proof of locality may arise from violating causal assumptions. Therefore, nonlocality is not always an instance of contextuality.

The requirement of no operational fine tuning that we develop in this work is strictly related to the notion of Leibnizianity introduced in \cite{Schmid2020unscrambling}. We argue that they represent the same concept: no operational fine tuning can be read as the requirement that the map from the ontological level (from the equivalences between conditional probabilities in the ontic extension, in our case) to the operational level (to the equivalences between conditional probabilities in the operational theory, in our case) must be one-to-one; Leibnizianity requires that the mapping from the operational level (from the ``inferential'' equivalences in the operational theory, in their case) to the ontological level (to the ``inferential'' equivalences in the ``unquotiented'' ontological model, in their case) must be one-to-one. The maps have opposite directions, but the concept is the same. Then, the main difference between the two works is the framework where these concepts are implemented. We use the framework of operational theories and ontic extensions, while \cite{Schmid2020unscrambling} uses the framework of causal-inferential theories, but, again, the concepts of no operational fine tuning and Leibnizianity capture the same idea. 
About the terminology, following a similar argument presented in \cite{Adlam2021}, we think that talking of no fine tuning is more appropriate than Leibnizianity as the reference to Leibniz's principle may be misleading. The property we are demanding for the map between the operational and ontological levels has its credentials in a methodological principle about how operationally equivalent scenarios should correspond to the same underlying reality, rather than a principle about the metaphysics of identity, like Leibniz's principle.\footnote{Let us mention that, interestingly, \cite{Adlam2021} also provides an explanation of the fine tunings associated to nonlocality and contextuality: they are a consequence of the impossibility of indeterministic closed causal loops within the all-at-once approach to physical theories \cite{Adlam2018}.} 

In this work we have related functors in category theory with operational theories and ontological models, which was already proposed and implemented by Gheorghiu and Heunen in \cite{Gheorghiu2019}. However, our approach differs from theirs both in the goal and in the mathematical formulation. Their goal was to address, in the categorical framework, the $\psi-$ontic versus $\psi-$epistemic issue on the reality of the quantum state \cite{LeiferOntology}, and, in order to do so, they defined operational categories in a much more structured way (also involving a notion of topology) than what we do here. 


The present work originates a proper research program, where the next step consists of formulating a resource theory of operational fine tunings.
This would also allow us to witness and quantify the presence of fine tunings in information processing tasks and quantum computational protocols.  As many results are showing \cite{ContextualityMBQC,Howard2014,Raussendorf2017,Delfosse2015,Vega2017,Delfosse2017,Catani2018,Henaut2018,Spekkens2009,vanDam, Barrett1,MansfieldKashefi2018,Schmid2018,Saha2019,SahaAnubhav2019,Yadavalli2020,LostaglioSenno2020}, the nonclassical phenomena proven to be responsible for the quantum computational advantages are so far dependent on the model and scenario considered, and this because, by construction, these phenomena arise only in certain setups. In this respect, the benefit of adopting the notion of fine tunings is that it captures the aspect that is common and inherently nonclassical about all such physical phenomena. Therefore, it may be possible that a certain amount of fine tuning, independent on which actual phenomenon is manifested in the setup considered, is necessary (or even sufficient!) for quantum computational advantages.  For these reasons, we believe that the notion of fine tunings is more promising than the ones so far explored in order to understand what powers quantum computers and technologies.  
Finally, the very foundational motivation for studying fine tunings, that represent a crucial problem in the interpretation and understanding of quantum theory, is to ultimately develop a new ontological framework for quantum theory absent of fine tunings, or, alternatively, explain them as emergent from yet undiscovered physical mechanisms. 

\section*{Acknowledgments}
The authors thank David Schmid and John Selby for helpful discussions. This research was supported by the Fetzer Franklin Fund of the John E. Fetzer Memorial Trust.


\bibliographystyle{unsrtnat}
\bibliography{FineTunings_biblio}

\begin{thebibliography}{71}
\providecommand{\natexlab}[1]{#1}
\providecommand{\url}[1]{\texttt{#1}}
\expandafter\ifx\csname urlstyle\endcsname\relax
  \providecommand{\doi}[1]{doi: #1}\else
  \providecommand{\doi}{doi: \begingroup \urlstyle{rm}\Url}\fi

\bibitem[Everett(1957)]{Everett1957}
Hugh Everett.
\newblock Relative state formulation of quantum mechanics.
\newblock \emph{Rev. Mod. Phys.}, 29:\penalty0 454--462, Jul 1957.
\newblock \doi{https://doi.org/10.1103/RevModPhys.29.454}.

\bibitem[Wallace(2012)]{Wallace2012}
David Wallace.
\newblock \emph{The Emergent Multiverse: Quantum Theory According to the
  Everett Interpretation}.
\newblock Oxford University Press, 2012.

\bibitem[Bohm(1952)]{Bohm1952}
David Bohm.
\newblock A suggested interpretation of the quantum theory in terms of "hidden"
  variables. i.
\newblock \emph{Phys. Rev.}, 85:\penalty0 166--179, Jan 1952.
\newblock \doi{https://doi.org/10.1103/PhysRev.85.166}.

\bibitem[D{\"u}rr and Teufel(2009)]{Durr2009}
Detlef D{\"u}rr and Stefan Teufel.
\newblock \emph{Bohmian Mechanics}, pages 145--171.
\newblock Springer Berlin Heidelberg, Berlin, Heidelberg, 2009.
\newblock \doi{https://doi.org/10.1007/b99978_8}.

\bibitem[Ghirardi et~al.(1986)Ghirardi, Rimini, and Weber]{Ghirardi1986}
G.~C. Ghirardi, A.~Rimini, and T.~Weber.
\newblock Unified dynamics for microscopic and macroscopic systems.
\newblock \emph{Phys. Rev. D}, 34:\penalty0 470--491, Jul 1986.
\newblock \doi{https://doi.org/10.1103/PhysRevD.34.470}.

\bibitem[Bassi et~al.(2013)Bassi, Lochan, Satin, Singh, and
  Ulbricht]{Bassi2013}
Angelo Bassi, Kinjalk Lochan, Seema Satin, Tejinder~P. Singh, and Hendrik
  Ulbricht.
\newblock Models of wave-function collapse, underlying theories, and
  experimental tests.
\newblock \emph{Rev. Mod. Phys.}, 85:\penalty0 471--527, Apr 2013.
\newblock \doi{https://doi.org/10.1103/RevModPhys.85.471}.

\bibitem[Rovelli(1996)]{Rovelli1996}
C.~Rovelli.
\newblock Relational quantum mechanics.
\newblock \emph{Int J Theor Phys}, 35:\penalty0 1637--1678, 1996.
\newblock \doi{https://doi.org/10.1007/BF02302261}.

\bibitem[Lombardi and Dieks(2017)]{Modalinterpretations}
Olimpia Lombardi and Dennis Dieks.
\newblock Modal interpretations of quantum mechanics.
\newblock In Edward~N. Zalta, editor, \emph{The Stanford Encyclopedia of
  Philosophy}. Metaphysics Research Lab, Stanford University, spring 2017
  edition, 2017.

\bibitem[Brukner and Zeilinger(2003)]{Brukner2003}
{\v{C}}aslav Brukner and Anton Zeilinger.
\newblock \emph{Information and Fundamental Elements of the Structure of
  Quantum Theory}, pages 323--354.
\newblock Springer Berlin Heidelberg, Berlin, Heidelberg, 2003.
\newblock ISBN 978-3-662-10557-3.
\newblock \doi{https://doi.org/10.1007/978-3-662-10557-3_21}.

\bibitem[Pitowsky(2006)]{Pitowsky2006}
Itamar Pitowsky.
\newblock \emph{Quantum Mechanics as a Theory of Probability}, pages 213--240.
\newblock Springer Netherlands, Dordrecht, 2006.
\newblock ISBN 978-1-4020-4876-0.
\newblock \doi{https://doi.org/10.1007/1-4020-4876-9_10}.

\bibitem[Fuchs et~al.(2014)Fuchs, Mermin, and Schack]{Fuchs2014}
Christopher~A. Fuchs, N.~David Mermin, and Rüdiger Schack.
\newblock An introduction to qbism with an application to the locality of
  quantum mechanics.
\newblock \emph{American Journal of Physics}, 82\penalty0 (8):\penalty0
  749--754, 2014.
\newblock \doi{https://doi.org/10.1119/1.4874855}.

\bibitem[Spekkens(2007)]{Spekkens2007}
Robert~W. Spekkens.
\newblock Evidence for the epistemic view of quantum states: A toy theory.
\newblock \emph{Phys. Rev. A}, 75:\penalty0 032110, Mar 2007.
\newblock \doi{https://doi.org/10.1103/PhysRevA.75.032110}.

\bibitem[Chiribella and Spekkens(2016)]{Spekkens2016}
Giulio Chiribella and Robert~W. Spekkens.
\newblock Quasi-quantization: classical statistical theories with an epistemic
  restriction.
\newblock In G.~Chiribella and R.~W. Spekkens, editors, \emph{Quantum Theory:
  Informational Foundations and Foils}, pages 1--20. Springer, Dordrecht, 2016.
\newblock URL \url{https://link.springer.com/book/10.1007/978-94-017-7303-4}.

\bibitem[Catani and Browne(2017)]{CataniBrowne2017}
Lorenzo Catani and Dan~E Browne.
\newblock Spekkens’ toy model in all dimensions and its relationship with
  stabiliser quantum mechanics.
\newblock \emph{New Journal of Physics}, 19\penalty0 (7):\penalty0 073035,
  2017.
\newblock \doi{https://doi.org/10.1088/1367-2630/aa781c}.

\bibitem[Catani et~al.(2021)Catani, Leifer, Schmid, and Spekkens]{Catani2021}
Lorenzo Catani, Matthew Leifer, David Schmid, and Robert~W. Spekkens.
\newblock Why interference phenomena do not capture the essence of quantum
  theory.
\newblock \emph{arXiv preprint arXiv:2111.13727}, 2021.
\newblock \doi{https://doi.org/10.48550/arxiv.2111.13727}.

\bibitem[Norsen(2017)]{Norsen2017}
Travis Norsen.
\newblock \emph{Foundations of Quantum Mechanics}.
\newblock Springer, first edition edition, 2017.
\newblock ISBN 978-3-319-65867-4.
\newblock \doi{https://doi.org/10.1007/978-3-319-65867-4}.

\bibitem[Bell(1966)]{Bell}
John~S. Bell.
\newblock On the problem of hidden variables in quantum mechanics.
\newblock \emph{Rev. Mod. Phys.}, 38:\penalty0 447--452, Jul 1966.
\newblock \doi{https://doi.org/10.1103/RevModPhys.38.447}.

\bibitem[Kochen and Specker(1967)]{KochenSpecker}
S.~Kochen and E.P. Specker.
\newblock The problem of hidden variables in quantum mechanics.
\newblock \emph{J. Math. Mech.}, 17:\penalty0 59--87, 1967.
\newblock \doi{http://doi.org/10.1512/iumj.1968.17.17004}.

\bibitem[Spekkens(2005)]{Spekkens2005}
R.~W. Spekkens.
\newblock Contextuality for preparations, transformations, and unsharp
  measurements.
\newblock \emph{Phys. Rev. A}, 71:\penalty0 052108, May 2005.
\newblock \doi{https://doi.org/10.1103/PhysRevA.71.052108}.

\bibitem[Price(2012)]{Price2012}
Huw Price.
\newblock Does time-symmetry imply retrocausality? how the quantum world says
  “maybe”?
\newblock \emph{Studies in History and Philosophy of Science Part B: Studies in
  History and Philosophy of Modern Physics}, 43\penalty0 (2):\penalty0 75 --
  83, 2012.
\newblock ISSN 1355-2198.
\newblock \doi{https://doi.org/10.1016/j.shpsb.2011.12.003}.

\bibitem[Leifer and Pusey(2017)]{LeiferPusey}
Matthew~S. Leifer and Matthew~F. Pusey.
\newblock Is a time symmetric interpretation of quantum theory possible without
  retrocausality?
\newblock \emph{Proceedings of the Royal Society A: Mathematical, Physical and
  Engineering Sciences}, 473\penalty0 (2202):\penalty0 20160607, 2017.
\newblock \doi{https://doi.org/10.1098/rspa.2016.0607}.

\bibitem[Leifer(2014)]{LeiferOntology}
Matthew Leifer.
\newblock Is the quantum state real? an extended review of psi-ontology
  theorems.
\newblock \emph{Quanta}, 3\penalty0 (1):\penalty0 67--155, 2014.
\newblock ISSN 1314-7374.
\newblock \doi{https://doi.org/10.12743/quanta.v3i1.22}.

\bibitem[Valentini(1996)]{Valentini1996}
Antony Valentini.
\newblock \emph{Pilot-Wave Theory of Fields, Gravitation and Cosmology}, pages
  45--66.
\newblock Springer Netherlands, Dordrecht, 1996.
\newblock \doi{https://doi.org/10.1007/978-94-015-8715-0_3}.

\bibitem[Weinberg(1989)]{Weinberg1989}
Steven Weinberg.
\newblock The cosmological constant problem.
\newblock \emph{Rev. Mod. Phys.}, 61:\penalty0 1--23, Jan 1989.
\newblock \doi{https://doi.org/10.1103/RevModPhys.61.1}.

\bibitem[Williams(2015)]{Williams2015}
Porter Williams.
\newblock Naturalness, the autonomy of scales, and the 125gev higgs.
\newblock \emph{Studies in History and Philosophy of Science Part B: Studies in
  History and Philosophy of Modern Physics}, 51:\penalty0 82--96, 2015.
\newblock ISSN 1355-2198.
\newblock \doi{https://doi.org/10.1016/j.shpsb.2015.05.003}.

\bibitem[Spekkens(2019)]{Spekkens2019}
Robert~W. Spekkens.
\newblock The ontological identity of empirical indiscernibles: Leibniz's
  methodological principle and its significance in the work of einstein.
\newblock \emph{arXiv.1909.04628'}, 2019.
\newblock \doi{https://doi.org/10.48550/arXiv.1909.04628}.

\bibitem[Pearl(2009)]{Pearl}
Judea Pearl.
\newblock \emph{Causality}.
\newblock Cambridge University Press, 2 edition, 2009.
\newblock \doi{https://doi.org/10.1017/CBO9780511803161}.

\bibitem[Wood and Spekkens(2015)]{Wood2015}
Christopher~J Wood and Robert~W Spekkens.
\newblock The lesson of causal discovery algorithms for quantum correlations:
  causal explanations of bell-inequality violations require fine-tuning.
\newblock \emph{New Journal of Physics}, 17\penalty0 (3):\penalty0 033002, mar
  2015.
\newblock \doi{https://doi.org/10.1088/1367-2630/17/3/033002}.

\bibitem[Harrigan and Spekkens(2010)]{Harrigan}
Nicholas Harrigan and Robert~W. Spekkens.
\newblock Einstein, {Incompleteness}, and the {Epistemic} {View} of {Quantum}
  {States}.
\newblock \emph{Foundations of Physics}, 40\penalty0 (2):\penalty0 125--157,
  2010.
\newblock \doi{https://doi.org/10.1007/s10701-009-9347-0}.

\bibitem[Leinster(2014)]{Leinster2014}
Tom Leinster.
\newblock \emph{Basic Category Theory}.
\newblock Cambridge Studies in Advanced Mathematics. Cambridge University
  Press, 2014.
\newblock \doi{https://doi.org/10.1017/CBO9781107360068}.

\bibitem[Jarrett(1984)]{Jarrett1984}
Jon~P. Jarrett.
\newblock On the physical significance of the locality conditions in the bell
  arguments.
\newblock \emph{No\^{u}s}, 18\penalty0 (4):\penalty0 569--589, 1984.
\newblock \doi{https://doi.org/10.2307/2214878}.

\bibitem[Ried et~al.(2015)Ried, Agnew, Vermeyden, Janzing, Spekkens, and
  Resch]{Ried2015}
Katja Ried, Megan Agnew, Lydia Vermeyden, Dominik Janzing, Robert~W. Spekkens,
  and Kevin~J. Resch.
\newblock A quantum advantage for inferring causal structure.
\newblock \emph{Nature Physics}, 11\penalty0 (5):\penalty0 414--420, May 2015.
\newblock ISSN 1745-2481.
\newblock \doi{https://doi.org/10.1038/nphys3266}.

\bibitem[Chaves et~al.(2015)Chaves, Majenz, and Gross]{Chaves2015}
Rafael Chaves, Christian Majenz, and David Gross.
\newblock Information--theoretic implications of quantum causal structures.
\newblock \emph{Nature Communications}, 6\penalty0 (1):\penalty0 5766, Jan
  2015.
\newblock ISSN 2041-1723.
\newblock \doi{https://doi.org/10.1038/ncomms6766}.

\bibitem[Fritz(2016)]{Fritz2016}
Tobias Fritz.
\newblock Beyond bell's theorem ii: Scenarios with arbitrary causal structure.
\newblock \emph{Communications in Mathematical Physics}, 341\penalty0
  (2):\penalty0 391--434, Jan 2016.
\newblock ISSN 1432-0916.
\newblock \doi{https://doi.org/10.1007/s00220-015-2495-5}.

\bibitem[Costa and Shrapnel(2016)]{Costa2016}
Fabio Costa and Sally Shrapnel.
\newblock Quantum causal modelling.
\newblock \emph{New Journal of Physics}, 18\penalty0 (6):\penalty0 063032, jun
  2016.
\newblock \doi{https://doi.org/10.1088/1367-2630/18/6/063032}.

\bibitem[Allen et~al.(2017)Allen, Barrett, Horsman, Lee, and
  Spekkens]{Allen2017}
John-Mark~A. Allen, Jonathan Barrett, Dominic~C. Horsman, Ciar\'an~M. Lee, and
  Robert~W. Spekkens.
\newblock Quantum common causes and quantum causal models.
\newblock \emph{Phys. Rev. X}, 7:\penalty0 031021, Jul 2017.
\newblock \doi{https://doi.org/10.1103/PhysRevX.7.031021}.

\bibitem[Weilenmann and Colbeck(2017)]{Weilenmann2017}
Mirjam Weilenmann and Roger Colbeck.
\newblock Analysing causal structures with entropy.
\newblock \emph{Proceedings of the Royal Society A: Mathematical, Physical and
  Engineering Sciences}, 473\penalty0 (2207):\penalty0 20170483, 2017.
\newblock \doi{https://doi.org/10.1098/rspa.2017.0483}.

\bibitem[Wolfe et~al.(01 Sep. 2019)Wolfe, Spekkens, and Fritz]{Wolfe2019}
Elie Wolfe, Robert~W. Spekkens, and Tobias Fritz.
\newblock The inflation technique for causal inference with latent variables.
\newblock \emph{Journal of Causal Inference}, 7\penalty0 (2):\penalty0
  20170020, 01 Sep. 2019.
\newblock \doi{https://doi.org/10.1515/jci-2017-0020}.

\bibitem[Vilasini and Colbeck(2019)]{Vilasini2019}
V.~Vilasini and Roger Colbeck.
\newblock Analyzing causal structures using tsallis entropies.
\newblock \emph{Phys. Rev. A}, 100:\penalty0 062108, Dec 2019.
\newblock \doi{https://doi.org/10.1103/PhysRevA.100.062108}.

\bibitem[Weilenmann and Colbeck(2020)]{Weilenmann2020}
Mirjam Weilenmann and Roger Colbeck.
\newblock Analysing causal structures in generalised probabilistic theories.
\newblock \emph{{Quantum}}, 4:\penalty0 236, February 2020.
\newblock ISSN 2521-327X.
\newblock \doi{https://doi.org/10.22331/q-2020-02-27-236}.

\bibitem[Barrett et~al.(2020)Barrett, Lorenz, and Oreshkov]{Barrett2020}
Jonathan Barrett, Robin Lorenz, and Ognyan Oreshkov.
\newblock Quantum causal models.
\newblock \emph{arXiv:1906.10726}, 2020.
\newblock \doi{https://doi.org/10.48550/arXiv.1906.10726}.

\bibitem[Cavalcanti(2018)]{Cavalcanti2018}
Eric~G. Cavalcanti.
\newblock Classical causal models for bell and kochen-specker inequality
  violations require fine-tuning.
\newblock \emph{Phys. Rev. X}, 8:\penalty0 021018, Apr 2018.
\newblock \doi{https://doi.org/10.1103/PhysRevX.8.021018}.

\bibitem[Landauer(1961)]{Landauer1961}
R.~Landauer.
\newblock Irreversibility and {Heat} {Generation} in the {Computing} {Process}.
\newblock \emph{IBM Journal of Research and Development}, 5\penalty0
  (3):\penalty0 183--191, 1961.
\newblock ISSN 0018-8646.
\newblock \doi{https://doi.org/10.1147/rd.53.0183}.

\bibitem[Minkowski()]{Minkowski1908}
H.~Minkowski.
\newblock Space and time - minkowski’s papers on relativity.
\newblock \emph{Qubec Canada: Minkowski Institute, reprinted in 2012}.

\bibitem[Goldstein et~al.(2002)Goldstein, Poole, and Safko]{Goldstein}
Herbert Goldstein, Charles~P. Poole, and John~L. Safko.
\newblock \emph{Classical Mechanics}.
\newblock Addison Wesley, third edition edition, 2002.
\newblock ISBN 0-201-65702-3.

\bibitem[Goldstein(2017)]{BohmianMechanics}
Sheldon Goldstein.
\newblock Bohmian mechanics.
\newblock In Edward~N. Zalta, editor, \emph{The Stanford Encyclopedia of
  Philosophy}. Metaphysics Research Lab, Stanford University, summer 2017
  edition, 2017.

\bibitem[Ghirardi(2018)]{CollapseTheories}
Giancarlo Ghirardi.
\newblock Collapse theories.
\newblock In Edward~N. Zalta, editor, \emph{The Stanford Encyclopedia of
  Philosophy}. Metaphysics Research Lab, Stanford University, fall 2018
  edition, 2018.

\bibitem[Cabello et~al.(2014)Cabello, Severini, and Winter]{CSW}
Adan Cabello, Simone Severini, and Andreas Winter.
\newblock Graph-{Theoretic} {Approach} to {Quantum} {Correlations}.
\newblock \emph{Phys. Rev. Lett.}, 112\penalty0 (4):\penalty0 040401, 2014.
\newblock \doi{https://doi.org/10.1103/PhysRevLett.112.040401}.

\bibitem[Ac{\'\i}n et~al.(2015)Ac{\'\i}n, Fritz, Leverrier, and
  Sainz]{Acin2015}
Antonio Ac{\'\i}n, Tobias Fritz, Anthony Leverrier, and Ana~Bel{\'e}n Sainz.
\newblock A combinatorial approach to nonlocality and contextuality.
\newblock \emph{Communications in Mathematical Physics}, 334\penalty0
  (2):\penalty0 533--628, 2015.
\newblock \doi{https://doi.org/10.1007/s00220-014-2260-1}.

\bibitem[Abramsky and Brandenburger(2011)]{Abramsky2011}
Samson Abramsky and Adam Brandenburger.
\newblock The sheaf-theoretic structure of non-locality and contextuality.
\newblock \emph{New Journal of Physics}, 13\penalty0 (11):\penalty0 113036, nov
  2011.
\newblock \doi{https://doi.org/10.1088/1367-2630/13/11/113036}.

\bibitem[Schmid et~al.(2020)Schmid, Selby, and
  Spekkens]{Schmid2020unscrambling}
David Schmid, John~H. Selby, and Robert~W. Spekkens.
\newblock Unscrambling the omelette of causation and inference: The framework
  of causal-inferential theories.
\newblock \emph{arXiv preprint arXiv:2009.03297}, 2020.
\newblock \doi{https://doi.org/10.48550/arXiv.2009.03297}.

\bibitem[Adlam(2021)]{Adlam2021}
Emily Adlam.
\newblock Contextuality, fine-tuning and teleological explanation.
\newblock \emph{Foundations of Physics}, 51\penalty0 (6):\penalty0 106, 2021.
\newblock \doi{https://doi.org/10.1007/s10701-021-00516-y}.

\bibitem[Adlam(2018)]{Adlam2018}
Emily Adlam.
\newblock Quantum mechanics and global determinism.
\newblock \emph{Quanta}, 7\penalty0 (1):\penalty0 40--53, 2018.
\newblock ISSN 1314-7374.
\newblock \doi{https://doi.org/10.12743/quanta.v7i1.76}.

\bibitem[Gheorghiu and Heunen(2020)]{Gheorghiu2019}
Alexandru Gheorghiu and Chris Heunen.
\newblock Ontological models for quantum theory as functors.
\newblock \emph{EPTCS}, 318:\penalty0 196--212, 2020.
\newblock \doi{https://doi.org/10.4204/EPTCS.318.12}.

\bibitem[Raussendorf(2013)]{ContextualityMBQC}
Robert Raussendorf.
\newblock Contextuality in measurement-based quantum computation.
\newblock \emph{Phys. Rev. A}, 88\penalty0 (2):\penalty0 022322, 2013.
\newblock \doi{https://doi.org/10.1103/PhysRevA.88.022322}.

\bibitem[Howard et~al.(2014)Howard, Wallman, Veitch, and Emerson]{Howard2014}
M.~Howard, J.~Wallman, V.~Veitch, and J.~Emerson.
\newblock Contextuality supplies the 'magic' for quantum computation.
\newblock \emph{Nature}, 510:\penalty0 351--355, 2014.
\newblock \doi{https://doi.org/10.1038/nature13460}.

\bibitem[Raussendorf et~al.(2017)Raussendorf, Browne, Delfosse, Okay, and
  Bermejo-Vega]{Raussendorf2017}
Robert Raussendorf, Dan~E. Browne, Nicolas Delfosse, Cihan Okay, and Juan
  Bermejo-Vega.
\newblock Contextuality and wigner-function negativity in qubit quantum
  computation.
\newblock \emph{Phys. Rev. A}, 95:\penalty0 052334, May 2017.
\newblock \doi{https://doi.org/10.1103/PhysRevA.95.052334}.

\bibitem[Delfosse et~al.(2015)Delfosse, Allard~Guerin, Bian, and
  Raussendorf]{Delfosse2015}
Nicolas Delfosse, Philippe Allard~Guerin, Jacob Bian, and Robert Raussendorf.
\newblock Wigner function negativity and contextuality in quantum computation
  on rebits.
\newblock \emph{Phys. Rev. X}, 5:\penalty0 021003, Apr 2015.
\newblock \doi{https://doi.org/10.1103/PhysRevX.5.021003}.

\bibitem[Bermejo-Vega et~al.(2017)Bermejo-Vega, Delfosse, Browne, Okay, and
  Raussendorf]{Vega2017}
Juan Bermejo-Vega, Nicolas Delfosse, Dan~E. Browne, Cihan Okay, and Robert
  Raussendorf.
\newblock Contextuality as a resource for models of quantum computation with
  qubits.
\newblock \emph{Phys. Rev. Lett.}, 119:\penalty0 120505, Sep 2017.
\newblock \doi{https://doi.org/10.1103/PhysRevLett.119.120505}.

\bibitem[Delfosse et~al.(2017)Delfosse, Okay, Bermejo-Vega, Browne, and
  Raussendorf]{Delfosse2017}
Nicolas Delfosse, Cihan Okay, Juan Bermejo-Vega, Dan~E. Browne, and Robert
  Raussendorf.
\newblock Equivalence between contextuality and negativity of the {Wigner}
  function for qudits.
\newblock \emph{New J. Phys.}, 19\penalty0 (12):\penalty0 123024, 2017.
\newblock ISSN 1367-2630.
\newblock \doi{https://doi.org/10.1088/1367-2630/aa8fe3}.

\bibitem[Catani and Browne(2018)]{Catani2018}
Lorenzo Catani and Dan~E. Browne.
\newblock State-injection schemes of quantum computation in spekkens' toy
  theory.
\newblock \emph{Phys. Rev. A}, 98:\penalty0 052108, Nov 2018.
\newblock \doi{https://doi.org/10.1103/PhysRevA.98.052108}.

\bibitem[Henaut et~al.(2018)Henaut, Catani, Browne, Mansfield, and
  Pappa]{Henaut2018}
Luciana Henaut, Lorenzo Catani, Dan~E. Browne, Shane Mansfield, and Anna Pappa.
\newblock Tsirelson's bound and landauer's principle in a single-system game.
\newblock \emph{Phys. Rev. A}, 98:\penalty0 060302, Dec 2018.
\newblock \doi{https://doi.org/10.1103/PhysRevA.98.060302}.

\bibitem[Spekkens et~al.(2009)Spekkens, Buzacott, Keehn, Toner, and
  Pryde]{Spekkens2009}
Robert~W. Spekkens, D.~H. Buzacott, A.~J. Keehn, Ben Toner, and G.~J. Pryde.
\newblock Preparation {Contextuality} {Powers} {Parity}-{Oblivious}
  {Multiplexing}.
\newblock \emph{Phys. Rev. Lett.}, 102\penalty0 (1):\penalty0 010401, 2009.
\newblock \doi{https://doi.org/10.1103/PhysRevLett.102.010401}.

\bibitem[van Dam(2000)]{vanDam}
B.~van Dam.
\newblock Nonlocality \& communication complexity.
\newblock \emph{PhD thesis, University of Oxford, Department of Physics}, 2000.

\bibitem[Barrett et~al.(2005)Barrett, Linden, Massar, Pironio, Popescu, and
  Roberts]{Barrett1}
Jonathan Barrett, Noah Linden, Serge Massar, Stefano Pironio, Sandu Popescu,
  and David Roberts.
\newblock Nonlocal correlations as an information-theoretic resource.
\newblock \emph{Phys. Rev. A}, 71\penalty0 (2):\penalty0 022101, 2005.
\newblock \doi{https://doi.org/10.1103/PhysRevA.71.022101}.

\bibitem[Mansfield and Kashefi(2018)]{MansfieldKashefi2018}
Shane Mansfield and Elham Kashefi.
\newblock Quantum advantage from sequential-transformation contextuality.
\newblock \emph{Phys. Rev. Lett.}, 121:\penalty0 230401, Dec 2018.
\newblock \doi{https://doi.org/10.1103/PhysRevLett.121.230401}.

\bibitem[Schmid and Spekkens(2018)]{Schmid2018}
David Schmid and Robert~W. Spekkens.
\newblock Contextual advantage for state discrimination.
\newblock \emph{Phys. Rev. X}, 8:\penalty0 011015, Feb 2018.
\newblock \doi{https://doi.org/10.1103/PhysRevX.8.011015}.

\bibitem[Saha et~al.(2019)Saha, Horodecki, and Paw{\l}owski]{Saha2019}
Debashis Saha, Pawe{\l} Horodecki, and Marcin Paw{\l}owski.
\newblock State independent contextuality advances one-way communication.
\newblock \emph{New Journal of Physics}, 21\penalty0 (9):\penalty0 093057, sep
  2019.
\newblock \doi{https://doi.org/10.1088/1367-2630/ab4149}.

\bibitem[Saha and Chaturvedi(2019)]{SahaAnubhav2019}
Debashis Saha and Anubhav Chaturvedi.
\newblock Preparation contextuality as an essential feature underlying quantum
  communication advantage.
\newblock \emph{Phys. Rev. A}, 100:\penalty0 022108, Aug 2019.
\newblock \doi{https://doi.org/10.1103/PhysRevA.100.022108}.

\bibitem[Yadavalli and Kunjwal(2022)]{Yadavalli2020}
Shiv~Akshar Yadavalli and Ravi Kunjwal.
\newblock Contextuality in entanglement-assisted one-shot classical
  communication.
\newblock \emph{{Quantum}}, 6:\penalty0 839, October 2022.
\newblock ISSN 2521-327X.
\newblock \doi{https://doi.org/10.22331/q-2022-10-13-839}.

\bibitem[Lostaglio and Senno(2020)]{LostaglioSenno2020}
Matteo Lostaglio and Gabriel Senno.
\newblock Contextual advantage for state-dependent cloning.
\newblock \emph{{Quantum}}, 4:\penalty0 258, April 2020.
\newblock ISSN 2521-327X.
\newblock \doi{https://doi.org/10.22331/q-2020-04-27-258}.

\end{thebibliography}

\end{document}